\documentclass[10pt]{IEEEtran}

\usepackage{color}
\usepackage{placeins}
\usepackage{float}
\usepackage[hidelinks]{hyperref}
\usepackage{tabularx,colortbl}
\usepackage[cmex10]{amsmath}
\usepackage{physics}
\usepackage{amssymb}
\usepackage[dvipsnames]{xcolor}

\usepackage[pdftex]{graphicx}
\graphicspath{{./FiguresforArxiv/}}
\usepackage[tight,footnotesize]{subfigure}

\def\d{\mathrm{d}}
\DeclareMathOperator{\sgn}{sgn}
\DeclareMathOperator{\sinc}{sinc}

\begin{document}
%
% paper title
% can use linebreaks \\ within to get better formatting as desired
%\title{Characterization and Design of Huygens' Metasurface Lenses for Extending the Angular Scan Range of Phased Arrays}
\title{Theory and Simulation of Metasurface Lenses for Extending the Angular Scan Range of Phased Arrays}

%% author names and affiliations
%% use a multiple column layout for up to three different
%% affiliations
%\author{\IEEEauthorblockN{Gleb A. Egorov}
%\IEEEauthorblockA{The Edward S. Rogers Department of Electrical \\ and Computer Engineering\\
%University of Toronto\\
%Toronto, Ontario, Canada\\
%Email: gleb.egorov@mail.utoronto.ca\\}
%\and
%\IEEEauthorblockN{George V. Eleftheriades}
%\IEEEauthorblockA{The Edward S. Rogers Department of Electrical \\ and Computer Engineering\\
%University of Toronto\\
%Toronto, Ontario, Canada\\
%Email: gelefth@waves.utoronto.ca}}
%%\and
%%\IEEEauthorblockN{Authors Name/s associated with 2nd Affiliation}
%%\IEEEauthorblockA{line 1: dept. name (if applicable)\\
%%line 2: name of organization, acronyms acceptable\\
%%line 3: City, State/Province, Country\\
%%line 4: e-mail address if desired}}

\author{Gleb~A.~Egorov, and~George~V.~Eleftheriades,~\IEEEmembership{Fellow,~IEEE}
%\thanks{Financial support from Huawei Canada is gratefully acknowledged.}
\thanks{G. A. Egorov is with the Edward S. Rogers Department of Electrical and Computer Engineering, University of Toronto,
Toronto, Ontario, Canada (e-mail: gleb.egorov@mail.utoronto.ca).}
\thanks{G. V. Eleftheriades is with the Edward S. Rogers Department of Electrical and Computer Engineering, University of Toronto,
Toronto, Ontario, Canada (e-mail: gelefth@waves.utoronto.ca).}}

% conference papers do not typically use \thanks and this command
% is locked out in conference mode. If really needed, such as for
% the acknowledgment of grants, issue a \IEEEoverridecommandlockouts
% after \documentclass

% for over three affiliations, or if they all won't fit within the width
% of the page, use this alternative format:
%
%\author{\IEEEauthorblockN{Michael Shell\IEEEauthorrefmark{1},
%Homer Simpson\IEEEauthorrefmark{2},
%James Kirk\IEEEauthorrefmark{3},
%Montgomery Scott\IEEEauthorrefmark{3} and
%Eldon Tyrell\IEEEauthorrefmark{4}}
%\IEEEauthorblockA{\IEEEauthorrefmark{1}School of Electrical and Computer Engineering\\
%Georgia Institute of Technology,
%Atlanta, Georgia 30332--0250\\ Email: see http://www.michaelshell.org/contact.html}
%\IEEEauthorblockA{\IEEEauthorrefmark{2}Twentieth Century Fox, Springfield, USA\\
%Email: homer@thesimpsons.com}
%\IEEEauthorblockA{\IEEEauthorrefmark{3}Starfleet Academy, San Francisco, California 96678-2391\\
%Telephone: (800) 555--1212, Fax: (888) 555--1212}
%\IEEEauthorblockA{\IEEEauthorrefmark{4}Tyrell Inc., 123 Replicant Street, Los Angeles, California 90210--4321}}

% use for special paper notices
%\IEEEspecialpapernotice{(Invited Paper)}

% make the title area
\maketitle

\begin{abstract}
Recent metasurface developments have led to their consideration for extending the angular scan range of phased arrays. This paper considers the use of Huygens' metasurfaces as lenses to accomplish the task. Ray optics is used to uncover how a single lens can extend the scan range. It is shown that a single scan extending lens leads to a broadside directivity degradation of the original beam. This directivity degradation is quantitatively characterized as a function of the desired angular scan-range expansion. A method of simulating phase boundaries is presented and used to verify the theoretical claims. A Huygens' metasurface lens design is presented and simulated to further validate the theoretical predictions and show that such lenses are physically realizable. %It is argued that using a general lens-like combination of surfaces to enhance array scan without reducing directivity at broadside is not possible. 
\end{abstract}


% For peer review papers, you can put extra information on the cover
% page as needed:
% \ifCLASSOPTIONpeerreview
% \begin{center} \bfseries EDICS Category: 3-BBND \end{center}
% \fi
%
% For peerreview papers, this IEEEtran command inserts a page break and
% creates the second title. It will be ignored for other modes.
\IEEEpeerreviewmaketitle

\section{Introduction}

Extending the scan performance of a given phased antenna array is a problem with a long history. Reference \cite{steyskal1979gain} was one of the first investigations of array scan enhancement. The work considered a theoretical phase-incurring dome which, when placed over the array, could steer the original $\pm60^\circ$-limited array beams all the way to the horizon. A physical dielectric dome was fabricated and tested in \cite{kawahara2007design}. Further theoretical and implementation advancements include design of refracting domes with transformation optics techniques \cite{lam2011steering,moccia2017transformation}, and reformulating the beam refraction problem as transferring gain pattern envelopes using power conservation \cite{kazim2013advanced,kazim2016wide}. All these phase incurring strucutres which achieved phased array scan enhancement did so at a cost of a degraded directivity, especially in the broadside direction. Furthermore, in \cite{abbaspour2013enhancing} it was observed that doubling the broadside directivity of the array via a single phase-incurring surface leads to halving of the scan range.

Recent developments in metasurface theory and design have reignited the community's interest in enhancing the scan range of antenna arrays \cite{silvestri2017dragonfly}. 
We will discuss the use of metasurface lenses, these are surfaces comprising sub-wavelength scatterers which act as conventional lenses, for this problem. In particular, it is possible to design a Huygens' metasurface to achieve a lossless and reflectionless transformation between the incident and transmitted fields \cite{epstein2016arbitrary}. Thus Huygens' metasurfaces can be made to provide a spatially varying phase shift, albeit perfectly reflectionless only with a prescribed excitation beam. Results of \cite{epstein2016arbitrary} will be used extensively throughout this publication and are summarized as follows. Let us say that we want a certain set of fields in one half space and a different set of fields in another half space. The boundary separating the two half spaces is the metasurface. The two sets of tangential fields on either side of the boundary are termed as postulated fields. If the two sets of fields locally conserve the real power crossing through the boundary, there exist generalized sheet transition conditions (GSTCs) which satisfy the boundary conditions of the two sets of fields. The GSTCs are given in the form of space-dependent impedance, admittance and magneto-electric coupling of the metasurface. The theoretical GSTCs can then be realized physically with unit cells composed of three metalization layers separated with dielectrics, which emulate Huygens' Omega bianisotropic unit cells \cite{chen2018theory}.

Due to the parallels between the existing theoretical work on phase-incurring surfaces and the metasurface-inspired surfaces in \cite{silvestri2017dragonfly} it is of no surprise that surface designs presented in \cite{silvestri2017dragonfly} also suffer from a degradation in broadside directivity. Reference \cite{benini2018phase} is another recent publication which considers the problem of array scan enhancement with phase-incurring metasurfaces. In that work a significant improvement in broadside directivity degradation was reported compared to existing literature. However, significant directivity degradation was reported elsewhere in the operating region of the device. Furthermore, the improvement at broadside was achieved with precise amplitude weighting and non-linear phasing of the antenna array elements. This approach is not necessarily compatible with all phased-array architectures, such as ones which use only phase control, clustering, random sub-arraying, overlapping or interleaved techniques \cite{rocca2016unconventional,azar2013overlapped}.

A natural question arising from the literature is whether it is possible to extend a phased-array scan range without incurring a drop in directivity and without resorting to unconventional array excitations. All the structures shown to achieve array scan enhancement in the listed publications resemble diverging lenses near broadside. References \cite{steyskal1979gain,kawahara2007design,kazim2013advanced,kazim2016wide,silvestri2017dragonfly} show that for a scan enhanced beam the worst directivity performance is observed around broadside. Thus, Sec. \ref{sec:ParaAnal} considers lens-like surfaces using ray optical theory. Moreover, the reason for why these scan-range extending surfaces resemble diverging lenses is presented. Directivity in the context of ray optics is also discussed and a quantitative relationship between directivity drop and angular scan range expansion is derived for the case of far-field lens placement. Finally, the possibility of placing lenses in the near field of the illuminating array is considered and broadside directivity is analysed.

In the process of answering the above question a need arose for a quick and simple simulation method to validate our reasoning. We have followed the common approach of studying scan enhancing metasurfaces, which is to treat them as phase-incurring boundaries. This was used in the context of wave optics as in \cite{steyskal1979gain}, and in conjunction with physical optics (see \cite{harrington1961time,gibson2007method} for description of the method) to compute scattering as in \cite{silvestri2017dragonfly,benini2018phase}. Sec. \ref{sec:PhaseBoundary} describes the MATLAB-implemented simulation method, and angle-doubling and angle-tripling phase boundaries were simulated to validate the derived directivity degradation of Sec. \ref{sec:DirTheory}. We note that the phase boundary view of metasurfaces is a rather simplistic one, and leads to some inconsistencies when it is combined with physical optics in the hope of obtaining reasonable electromagnetic behavior. These inconsistencies, and when they are most pronounced, are discussed in App. \ref{sec:AppPhaseCloserLook}.

Finally, in Sec. \ref{sec:MetaLens}, a Huygens' Omega bianisotropic lens is designed and full-field simulated to verify the phase surface simulation results and to show that such unconventional lenses are feasible. The full-field simulation is performed in COMSOL Multiphysics, and the metasurface is implemented with ideal impedance sheets.

Due to computational resource limitations, all analysis and simulations presented in this publication assume the problems to be two dimensional, i.e. the lens/metasurface parameters and incident fields/beams are functions of only two spatial variables. Furthermore, only TE-polarized fields are considered.

\section{Theory of Scan Enhancement} \label{sec:ParaAnal}

\subsection{Diverging Lens for Scan Enhancement} \label{sec:LensEnh}

\begin{figure}[t!]
\begin{center}
\noindent
  \subfigure[]{\includegraphics[width=3in]{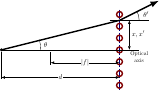}}
  \subfigure[]{\includegraphics[width=3in]{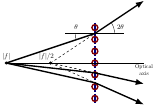}}
  \caption{Ray digrams for a lens-like metasurface. The surface is depicted as an array of electric and magnetic dipoles, which incur a desired phase onto incident fields. (a) Depicts directions and positions at the input and output faces of the metasurface for a ray passing through the device. (b) Depiction of angular doubling. By placing a ray source at the focal point of a diverging lens, all rays double their angle as they pass through the device. The outgoing rays appear to emanate from a position $|f|/2$ behind the lens. %A beam of angular spread $\delta$ has a spread of $2\delta$ after passing through the lens. }\label{fig:rays}
  }\label{fig:rays}
\end{center}
\end{figure}

A single diverging lens can provide an arbitrary scan enhancement to array beams in the context of paraxial ray optics. This can be understood by considering what happens to a ray which originates on the optical axis at a distance $d$ away from the lens of focal length $f$, as shown in Fig. \ref{fig:rays}a. The ray transfer matrix for a lens is employed to find the direction and position of the ray once it passes through the lens \cite{yariv2007photonics}:
\begin{equation} \label{eq:matrix}
\begin{bmatrix}
x' \\
\theta'
\end{bmatrix} 
= 
\begin{bmatrix}
1 & 0 \\
-1/f & 1
\end{bmatrix}
\begin{bmatrix}
\theta d \\
\theta
\end{bmatrix}
\end{equation}
The quantities appearing in the above equation are depicted in Fig. \ref{fig:rays}a. From (\ref{eq:matrix}) we obtain
\begin{equation} \label{eq:theta2}
\theta' = \alpha\theta = \left(1-\frac{d}{f}\right)\theta,
\end{equation}
where $\alpha$ is defined as the scan enhancement factor.

%Equation (\ref{eq:theta2}) shows that an adequate choice of $z$ and $f$ leads to an arbitrary angle enhancement of an incident beam. For example placing a ray source at the focal point of a concave lens (value of $f$ is negative by convention) leads to an $\alpha$ of 2. In this case the lens places the image of the source at half the focal length which leads to the doubling of ray angles. This behavior is depicted in Fig. \ref{fig:rays}b. An antenna array placed at $z$ provides rays emanating approximately from the optical axis. The array beam is centered at $\theta_0$ with an angular spread of $\delta$. It is easy to see that once the beam passes through the lens, it points in the $\alpha\theta_0$ direction while having a spread of $\alpha\delta$. This shows that the directivity of the beam degrades with enhanced scan. This behavior is depicted for the case of $\alpha=2$ in Fig. \ref{fig:rays}b.

Equation (\ref{eq:theta2}) shows that an adequate choice of $d$ and $f$ leads to an arbitrary angle enhancement of an incident beam. For example, placing a ray source at the focal point of a concave lens (value of $f$ is negative by convention) leads to an $\alpha$ of 2. In this case, the lens places the image of the source at half the focal length, which leads to the doubling of ray angles. This behavior is depicted in Fig. \ref{fig:rays}b. An antenna array placed at $d$ provides rays emanating approximately from the optical axis. The rays constitute a beam pointing in a given direction. As the rays pass through the lens, they acquire a common angular enhancement. Thus the beam changes direction but also acquires a different angular spread, which leads to a change in directivity. %The array beam is centered at $\theta_0$ with an angular spread of $\delta$. It is easy to see that once the beam passes through the lens, it points in the $\alpha\theta_0$ direction while having a spread of $\alpha\delta$. This shows that the directivity of the beam degrades with enhanced scan. This behavior is depicted for the case of $\alpha=2$ in Fig. \ref{fig:rays}b.

\subsection{Effect of Scan Enhancement on Directivity} \label{sec:DirTheory}

In the context of ray optics, a beam is composed of a collection of $N$ rays. Each ray has a direction $\theta$ in which it propagates. It is possible to define a density of rays as a function of direction, $\rho(\theta)$. For any beam, which is to be used with a lens for scan enhancement, the number of rays propagating in the angular range from $\theta$ to $\theta{+}\d\theta$ is equal to $\d N {=} \rho(\theta)\d\theta$. Furthermore, it can be thought that a single ray carries a nominal amount of power. Thus the total power in a beam is proportional to the total number of rays, which in turn is given by $N {=} \int\rho(\theta)\d\theta$.

We define the ray optical directivity as
\begin{equation} \label{eq:2DDirRho}
D(\theta) = \frac{2\pi\rho(\theta)}{\int\rho(\beta)\d\beta}.
\end{equation}
This definition is discussed further in App. \ref{sec:AppRayD2D}.
%This definition is limited to the case where the studied beam is in the far field of the source, such that the ray optical viewpoint describes it well. For example applying this definition of directivity to a Gaussian beam at its waist leads to a delta function since there all rays are exactly colinear. However, such a value of directivity is unrealistic as the beam diverges away from the waist. 

\begin{figure}[t!]
\begin{center}
\includegraphics[width=3.5in]{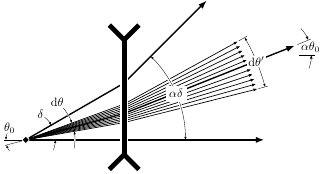}
\caption{A beam incident on a diverging lens. The center of the beam travels at $\theta_0$ and has an angular spread of $\delta$. After passing through the lens, the beam points in the $\alpha\theta_0$ direction and has an angular spread of $\alpha\delta$. A small bundle of rays centered around $\theta_0$ is depicted. The bundle has an angular spread of $\mathrm{d}\theta$. As the bundle passes through the lens, it spans $\mathrm{d}\theta'$ around $\alpha\theta_0$. Note that the number of rays in the bundle stays constant as it passes through the lens.
}\label{fig:directivity}
\end{center}
\end{figure}

To see how the directivity of a beam passing through a scan enhancing lens is affected, consider what happens to a small bundle of rays as shown in Fig. \ref{fig:directivity}. There, a beam is propagating in the direction $\theta_0$. Each ray experiences a scan enhancement of $\alpha$. Therefore, the scan enhanced beam points in the $\theta'_0=\alpha\theta_0$ direction. The scan enhanced beam has a different angular distribution of rays, quantified by the ray density function $\rho'(\theta')$. The bundle of rays of Fig. \ref{fig:directivity} contains $\d N = \rho(\theta)\d\theta$ rays. As this bundle passes through the lens, the number of rays remains unchanged due to energy conservation. Thus $\d N = \rho(\theta)\d\theta=\rho'(\theta')\d\theta'$. Using the relation $\theta'=\alpha\theta$ it is easy to see that $\rho'(\alpha\theta) = \rho(\theta)/\alpha$. Again due to energy conservation $\int\rho'(\theta')\d\theta'=\int\rho(\theta)\d\theta$, which is used to show that
\begin{equation} \label{eq:DirPerfAlpha}
\frac{D'(\alpha\theta)}{D(\theta)}=\frac{1}{\alpha}.
\end{equation}
Therefore a scan enhancing setup with $\alpha$ of 2 would lead to halving the original beam directivity, which is in agreement with past work \cite{kazim2013advanced, kazim2016wide, abbaspour2013enhancing}. 

\subsection{Scan Enhancement of Finitely Sized Arrays} \label{sec:FiniteArray}

Up to now the discussion of scan enhancement on directivity has been limited to large distances between the source and scan enhancing lens. However, it is generally desired to place a scan enhancing structure as close as possible to its source. In this section, we consider what happens when an angle-doubling lens ($\alpha{=}2$ in (\ref{eq:theta2}), which makes $d{=}{-}f$) is placed close enough such that the spatial extent of the illuminating source cannot be ignored. Furthermore, only the case of a broadside incident beam is considered in this section.

For this discussion, another measure of directivity is useful. It is shown in App. \ref{sec:AppDUmax} that the radiation pattern of a uniform two-dimensional aperture of length $L$ has its peak in the broadside direction, with peak directivity of
\begin{equation} \label{eq:DirUmax}
D_{U,\mathrm{max}}=\frac{2\pi L}{\lambda}.
\end{equation}
When the source is not a uniform aperture but a uniformly excited antenna array, this measure of peak directivity is still valid as long as no grating lobes are present in the visible region of the array (i.e. array element spacing is less than $\lambda/2$). Note that the phrase ``uniformly excited array" implies the antenna elements are oscillating with the same magnitude and phase.

\begin{figure}[t!]
\begin{center}
\includegraphics[width=3.5in]{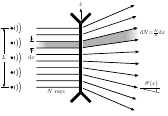}
\caption{Illumination of an angle-doubling lens by a uniformly excited array of length $L$. The array produces a collimated beam, composed of $N$ uniformly distributed rays, which diverges after the lens. A ray which reaches the lens at $x$ is refracted towards $\theta'(x)$. An infinitesimal section $\mathrm{d}x$ of the incident beam carries $\mathrm{d}N$ rays, which is conserved while passing through the lens.
}\label{fig:UnifArr}
\end{center}
\end{figure}

With the angle-doubling lens being in the vicinity of the array, the incident rays are collimated in the broadside direction. These incident rays are assumed to be uniformly distributed in a beam of width $L$. The lens diverges these rays as if they originated on the optic axis a distance $|f|$ behind the lens. This scenario is shown in Fig. \ref{fig:UnifArr}, along with some quantities useful for directivity analysis. Using this problem setup and following the reasoning of Sec. \ref{sec:DirTheory}, one can find the resulting beam directivity to be
\begin{equation} \label{eq:DirSec^2}
D'(\theta')=
\begin{cases}
	\frac{2\pi |f|}{L} \sec^2\theta',& \text{if } |\theta'| \leq \theta'_M\\
	0, & \text{otherwise},
\end{cases}
\end{equation}
where $\theta'_M{=}\atan \frac{L}{2|f|}$. It is not expected that the refracted beam directivity would exhibit discontinuities at $\theta'{=}\pm\theta'_M$, which appear simply because the incident beam is assumed to be discontinuous. Nevertheless, the obtained expression is expected to be accurate away from the two discontinuities and lens performance can be assessed.  First of all, this shows that the refracted beam no longer exhibits peak directivity in the desired broadside direction. Because (\ref{eq:DirUmax}) and (\ref{eq:DirSec^2}) are valid for large $L$ and small $|f|$, it is easy to notice that $D'(0)/D_{U,\mathrm{max}}$ can be much smaller than $1/2$, implying possible broadside directivity degradation far greater than the 3dB degradation one would have obtained by placing the lens in the far-field of the source.

\begin{figure}[t!]
\begin{center}
\includegraphics[width=3.5in]{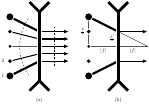}
\caption{Obtaining directive radiation from a closely-placed angle-doubling lens. The size of the point sources and rays emanating from them represent the relative magnitude of their excitation. (a) shows the array excited in a non-uniform fashion to produce a converging beam, which in turn is refracted by the lens into a collimated one. The thin dashed lines represent surfaces of constant phase of the input and output beams. Non-uniform phasing of the the source array is obviously required to obtain the desired curved phase-front. (b) shows the outermost rays of the beam and the two triangles which are used to relate the length of the array and the output aperture.
}\label{fig:MaciPhasing}
\end{center}
\end{figure}

This shows that placing an angle-doubling lens close to a uniformly excited array does not lead to directive radiation in the desired broadside direction, but instead peak directivity is obtained in the $\pm\theta_M$ directions. Nevertheless, it is possible to achieve directive broadside radiation from a closely placed angle-doubling lens. To do so, the phased array elements have to be excited in a non-uniform fashion. The required excitation method is described in detail in \cite{benini2018phase}. This method phases the antenna elements in a way which creates a converging beam, which after passing through the angle-doubling lens becomes uniform in phase. Such element phasing (``pre-distortion") essentially cancels the phase of the lens at its output. At the same time, the excitation magnitude of the antenna elements is also non-uniform and is chosen in a way which causes the resulting output aperture to be uniform in magnitude. This scenario is depicted in Fig. \ref{fig:MaciPhasing}(a). Thus, in order to analyze broadside directivity for this array excitation one simply needs to compare the aperture length $L'$ on the transmission side of the lens to the length of the array $L$.

The output aperture length $L'$ is dictated by the refraction which has to take place to produce a collimated beam of flat phase on the transmission side of the lens. To obtain $L'$, one has to consider the path of a ray from one of the edge array elements. This ray has to travel in the direction of the optic axis once refracted by the lens. This occurs when the ray initially travels towards the focal point on the transmission side of the lens. From there it is easy to show using similar triangles that for an angle-doubling lens, $L'{=}L/2$. The paths of the outermost rays, along with the dimensions of the relevant similar triangles are shown in Fig. \ref{fig:MaciPhasing}(b). According to (\ref{eq:DirUmax}) this means that a 3dB degradation in broadside directivity is still present.

Although it appears that this method allows one to place an angle-doubling lens arbitrarily close to the source array while still suffering only a 3dB directivity degradation at broadside, it is not always the case. To demonstrate this, consider the case when the element spacing, $d_e$, is $\lambda/2{<}d_e{<}\lambda$. When phased in a linear fashion, such an array would start producing a grating lobe peak when the main beam is scanned off broadside to \cite{balanis2016antenna}
\begin{equation}
\theta_g{=}\arcsin\left(\frac{\lambda}{d_e}-1\right).
\end{equation}
We now place an angle-doubling lens in the vicinity of the array to double its grating-lobe-free scan range. As described above, in order to obtain a uniform aperture on the transmission side of the lens, the array must be phased non-linearly, with the beam focusing on the optic axis a distance $2|f|$ away from the array (see Fig. \ref{fig:MaciPhasing}b). Such phasing causes larger interelement phasing for the outer elements, with the largest phase difference occurring for the two adjacent outermost elements on both sides of the array. We now consider the bottom two elements (elements labeled 1 and 2 in Fig. \ref{fig:MaciPhasing}a) as a two-element array. In order to contribute to the uniform output aperture this two-element array must produce a beam in the direction 
\begin{equation}
\theta_a = \arctan\left(\frac{L-d_e}{4|f|}\right).
\end{equation}
If $\theta_a \geq \theta_g$, the two-element array will exhibit a grating lobe peak and the directivity of the desired beam will be reduced. This condition occurs when 
\begin{equation} \label{eq:Lmaxcond}
L \geq 4|f|\frac{\lambda-d_e}{\sqrt{2\lambda d_e-\lambda^2}} + d_e.
\end{equation}
Note that for $L$ significantly larger than the right hand side of the above expression, not only the outermost two elements produce a grating lobe, but other elements away from the edges start contributing to radiation in undesired directions. This further reduces the directivity performance of the angle doubler. Also note that this reasoning does not lead to a measure of the directivity reduction and is meant only as a rule of thumb. Nevertheless, we can say that in order for an angle-doubling lens to maintain a 3dB directivity drop at broadside, the array cannot be larger in length than the right hand side of (\ref{eq:Lmaxcond}).

\section{Electromagnetic Phase Surfaces} \label{sec:PhaseBoundary}

In order to allow for the electromagnetic simulation of electrically large problems, we consider a simplified view of metasurfaces -- namely the view that metasurfaces can be treated as phase incurring boundaries. A metasurface provides a transition between the tangential fields on its two sides via electric and magnetic currents \cite{epstein2016arbitrary,kuester2003averaged}. Because it is the only quantity describing the surface, the phase function must provide a way to obtain these tangential fields. Furthermore, we restrict our attention only to lossless and passive surfaces. This implies that for any point on the surface the real power passing though it has to be conserved \cite{epstein2016arbitrary}. 

\begin{figure}[t!]
\begin{center}
\noindent
  \includegraphics[width=3.5in]{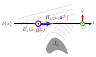}
  \caption{Geometry used during phase boundary definition. The phase boundary itself is shown by the black line. The boundary is illuminated by a general current distibution which produces TE fields. The tangential incident and transmitted $E_z$ and $H_x$ fields at a point on the boundary are depicted with the red-blue vectors.}\label{fig:PhaseSurfGeom}
\end{center}
\end{figure}

With this in mind, the operation of a phase boundary can be defined in an obvious fashion as follows. Sources below the surface illuminate it and establish incident tangential electric and magnetic fields on it. These incident tangential fields are taken as the total tangential fields on the incident side of the surface, namely $E_z(x,y{=}0^-)$ and $H_x(x,y{=}0^-)$ for TE polarization. Note that the explicit $y$-dependence is shown here for clarity and will be subsequently omitted. The considered geometry is shown in Fig. \ref{fig:PhaseSurfGeom}. The figure depicts the coordinate system used throughout this publication, along with the phase boundary lying in the $y{=}0$ plane, which is illuminated by a general $z$-directed current distribution lying below the boundary. The tangential fields on the transmitted side are calculated as
\begin{align} 
E_z(x,0^+) &= E_z(x,0^-)e^{\mathrm{j}\phi(x)}, \label{eq:PhaseBoundaryE}\\
H_x(x,0^+) &= H_x(x,0^-)e^{\mathrm{j}\phi(x)}. \label{eq:PhaseBoundaryH}
\end{align}
Note that indeed these tangential fields satisfy power flow conservation, since the component of the Poynting vector normal to the surface is conserved across the boundary:  $E_z(x,0^-)H_x^*(x,0^-){=}E_z(x,0^+)H_x^*(x,0^+)$. According to \cite{epstein2016arbitrary}, since power flow through the surface is conserved there exists a physical Omega bianisotropic surface which achieves this field transformation. It is worth emphasizing the generality of the presented phase boundary definition. This field transformation is written in a way which can be used with arbitrary incident fields. Furthermore, a single phase boundary surface can be subject to various incidence scenarios, which is useful for simulating metasurface behavior subject to incidence from a steerable antenna array.

One issue with attempting to realise such boundary with a physical metasurface is immediately apparent. This phase surface can be subject to various incidence cases. An Omega bianisotrpic surface, which is defined with a specific incidence in mind, will not in general behave in the same fashion when subject to different incidence fields. Thus, strictly speaking, using a single phase boundary with many incidence scenarios does not mean that a single physical surface achieves this performance. Instead, the different incidence cases would result in different Omega bianisotropic surfaces which produce the same phase discontinuity. Nevertheless, one can disregard this complication of using phase boundaries and still simulate the scattering by such a fictitious surface as was done in \cite{silvestri2017dragonfly,benini2018phase}. After all, as Sec. \ref{sec:MetaLens} will show, this approach has some merit when the different incidence fields correspond to near-broadside incident beams. This is not the only inconsistency that this simulation method suffers from and a user must be well informed of them to be certain in validity of simulation results. Other notable shortcomings are discussed in App. \ref{sec:AppPhaseCloserLook}. However, it must be noted that this method does not require a tedious physical surface design and is significantly less resource intensive compared to full-field metasurface simulations. 

Having defined the concept of a phase boundary, the next natural question is how to simulate the scattering produced by such an interface? One straight-forward approach is to use the tangential fields on the transmission side of the surface, employ Love's equivalence principle to these transmitted fields to obtain surface electric and magnetic currents. These surface currents are discretized and radiation from individual elements can be easily computed via well known formulas \cite{gibson2007method}. 

We employ the equivalence principle on a closed boundary which surrounds the antenna array and the surface. Also, part of this boundary comes infinitesimally close to the transmission side of the phase surface. This section of the boundary is given the tangential transmitted phase surface fields. For the rest of the boundary, the tangential fields correspond to the ones produced by the illuminating array. The equivalence principle is then used in such a way that sets the fields inside this boundary to 0, while maintaining the same radiated fields \cite{harrington1961time}. The boundary on which the equivalence principle is applied will be referred to as the equivalence box. The equivalence box is depicted in Fig. \ref{fig:PhaseSurfArrEquiv}, which encompasses the phase boundary and its source.

\subsection{Simulating Angle-Doubling Phase Surfaces} \label{sec:SimPhaseSurf}

\begin{figure}[t!]
\begin{center}
\noindent
  \includegraphics[width=3.5in]{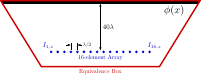}
  \caption{Geometry of the simulated angle-doubling phase boundary. The 16-element array of infinitely long $z$-directed current lines and the equivalence box are shown. Drawn as an illustration and not to scale.}\label{fig:PhaseSurfArrEquiv}
\end{center}
\end{figure}

The described simulation method was programmed in MATLAB, and an angle-doubling lens was simulated, providing the opportunity to verify and discuss the theory presented in the previous section. In this simulation, the incident beam is produced by a uniformly excited, 16 element phased array with a half-wavelength element spacing. Array elements are infinite lines of current, extending in the $z$-direction, which allow for the two-dimensional treatment of the problem. Due to the source configuration only TE fields are established. The array and associated geometry are shown in Fig. \ref{fig:PhaseSurfArrEquiv}. As previously, the metasurface lies in the $y{=}0$ plane. The incident beam is limited to a scan range of $\pm15^\circ$ off broadside. It is desired for the surface to extend the scan range of the phased array to $\pm30^\circ$. %The frequency of operation is chosen to be 10 GHz.
This section attempts to validate (\ref{eq:DirPerfAlpha}), which assumes that the lens is located in the far field of the source and thus we chose a large value. The distance between the phased array and the lens is chosen to be 40$\lambda$. Although this distance is too large to be useful practically at conventional microwave frequencies, it is not the case for the millimeter-wave spectrum which is currently considered for 5G communications \cite{rappaport2017overview,liu2017will}. A metasurface lens of focal length $f$ exhibits the following phase \cite{elsakka2016multifunctional}
\begin{equation} \label{eq:LensPhase}
\phi(x)=\sgn(f) k \sqrt{x^2 + f^2}-kf,
\end{equation}
where $\sgn(\cdot)$ is the sign function. This phase function was implemented and setting $f{=}{-}40\lambda$ led to an angle-doubling phase boundary.

\begin{figure}[t!]
\begin{center}
\noindent
  \subfigure[]{\includegraphics[width=3.5in]{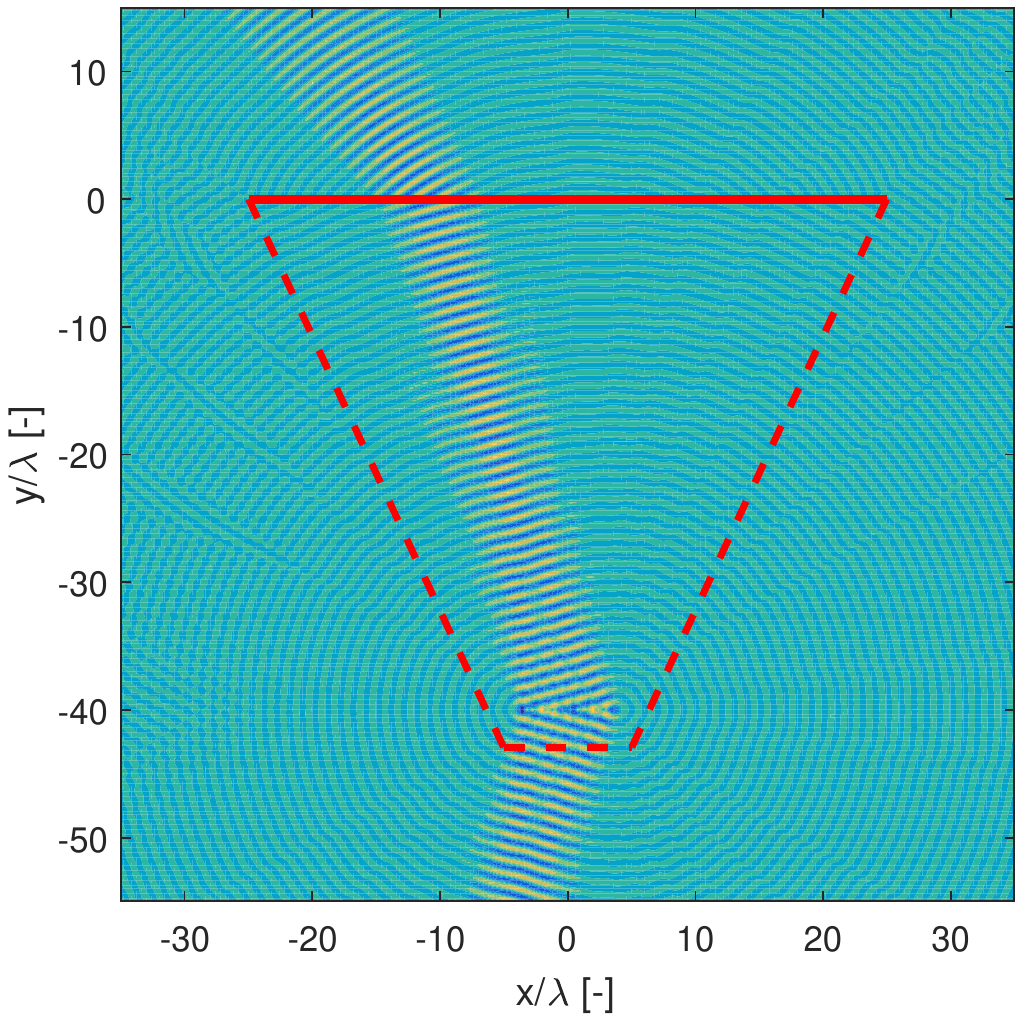}}
  \subfigure[]{\includegraphics[width=3.5in]{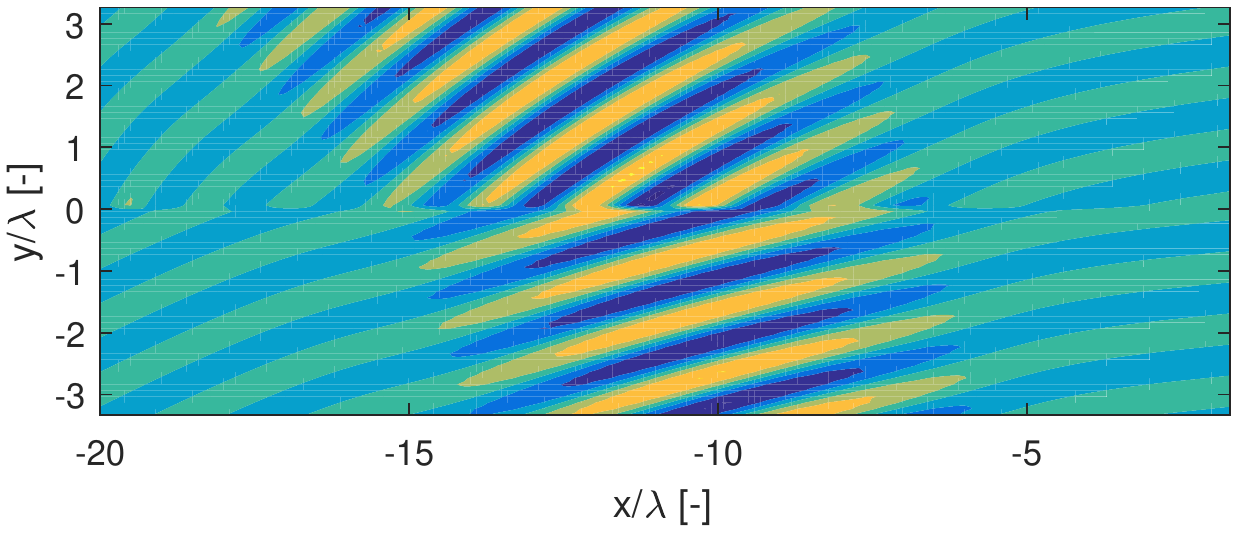}}
  \caption{Simulated behavior of the angle-doubling phase boundary. The out-of-plane electric field is plotted. (a) shows the array illuminating the phase surface with a 15$^\circ$ off broadside beam, which is refracted to 30$^\circ$. The 16 element array is located in the $y{=}{-}40\lambda$ plane. The red lines show the location of the equivalence box. The solid red line coincides with the location of the phase boundary. (b) zooms in on the refraction taking place at the surface.
  }\label{fig:PhaseSim15to30}
\end{center}
\end{figure}

\begin{figure}[t!]
\begin{center}
\noindent
  \includegraphics[width=3.5in]{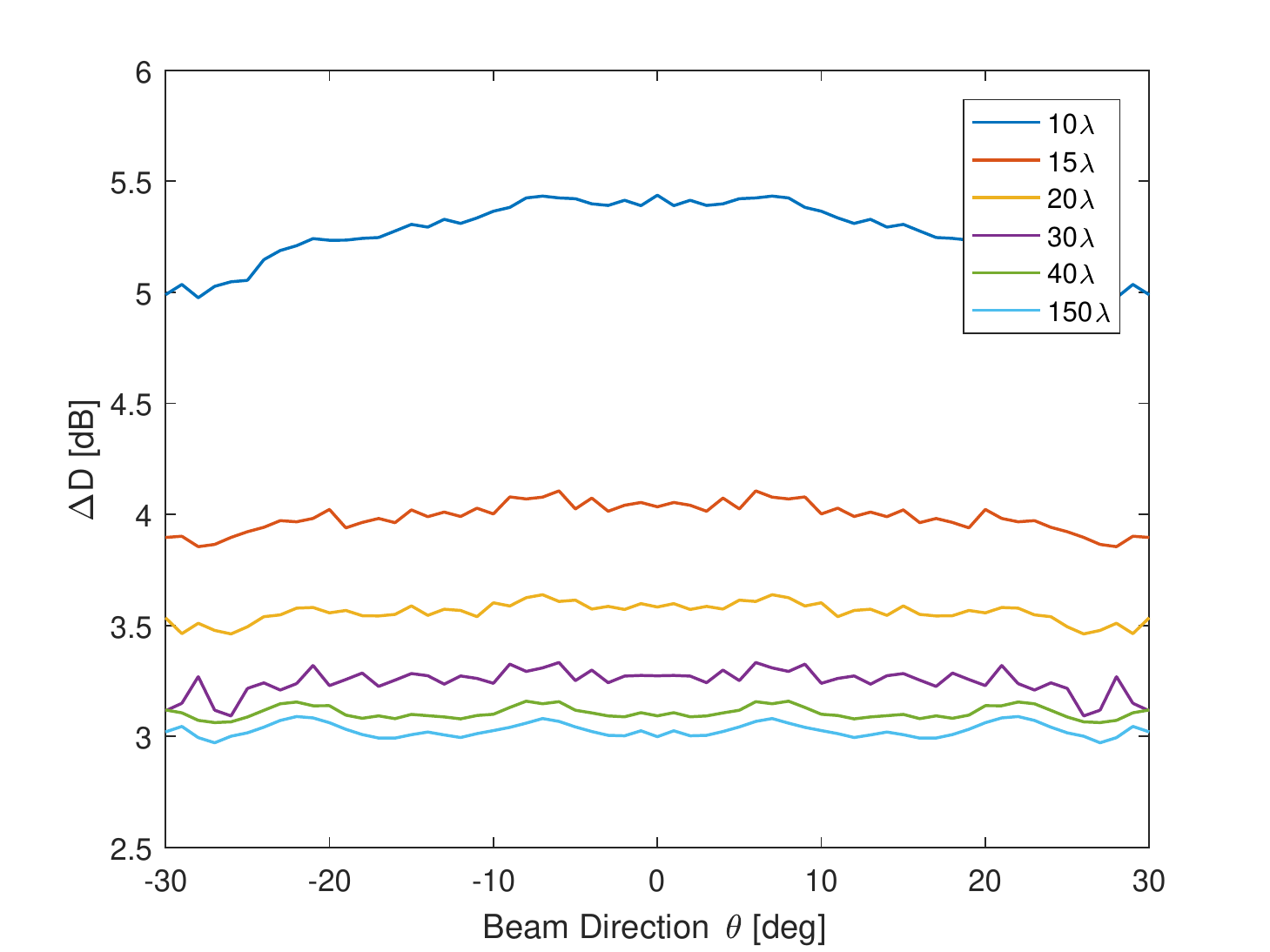}
  \caption{The directivity difference between the array itself and the scan extending lens at the same beam angle. Each curve corresponds to a different angle doubling scenario -- different lenses and array-lens separations. The legend gives the array-lens distance for each curve.}\label{fig:AngleDoublerAlphavsD}
\end{center}
\end{figure}

Fig. \ref{fig:PhaseSim15to30} depicts the case where the array produces a 15$^\circ$ off broadside beam. Indeed a 15$^\circ$ to 30$^\circ$ refraction is observed. The equivalence box is depicted with red solid and dashed lines. The solid line corresponds to the location of the phase boundary. For this and future field plots of phase surface simulation data, the fields inside the equivalence box are simply the radiated fields by the array within it. Also note that for illustrative purposes, the length of the lens appearing in Fig. \ref{fig:PhaseSim15to30} is significantly smaller than what was used to obtain the performance data of Figs. \ref{fig:PhaseSimDirPerfPhaseSurf}, \ref{fig:AngleDoublerAlphavsD}, \ref{fig:AngleTripDirPerf} and \ref{fig:AngleTripDirComp}. There, the lenses were chosen to be 300$\lambda$ long.

Fig. \ref{fig:PhaseSimDirPerfPhaseSurf} depicts the scanning performance of the angle-doubling phase boundary. The performance is plotted versus the incident beam angle. The $\theta_r$ of the right ordinate represents the angle at which the refracted beam has maximum directivity. The left ordinate depicts the directivity performance of the refracted beams. $D_\mathrm{max}$ is the maximum value of a refracted beam with a given excitation. $D(2 \theta_i)$ is the value of directivity of the refracted beam in the direction of $2\theta_i$. It is observed that the device approximates an angle doubler extremely well. The error in output beam angles is within $\pm1^\circ$ compared to a perfect doubler. The discrepancy between peak directivity and the obtained directivity at double of the incident angle is insignificant, and occurs only at the edges of the operating region of the device.

{\color{Red}
\begin{figure}[t!]
\begin{center}
\noindent
  \includegraphics[width=3.5in]{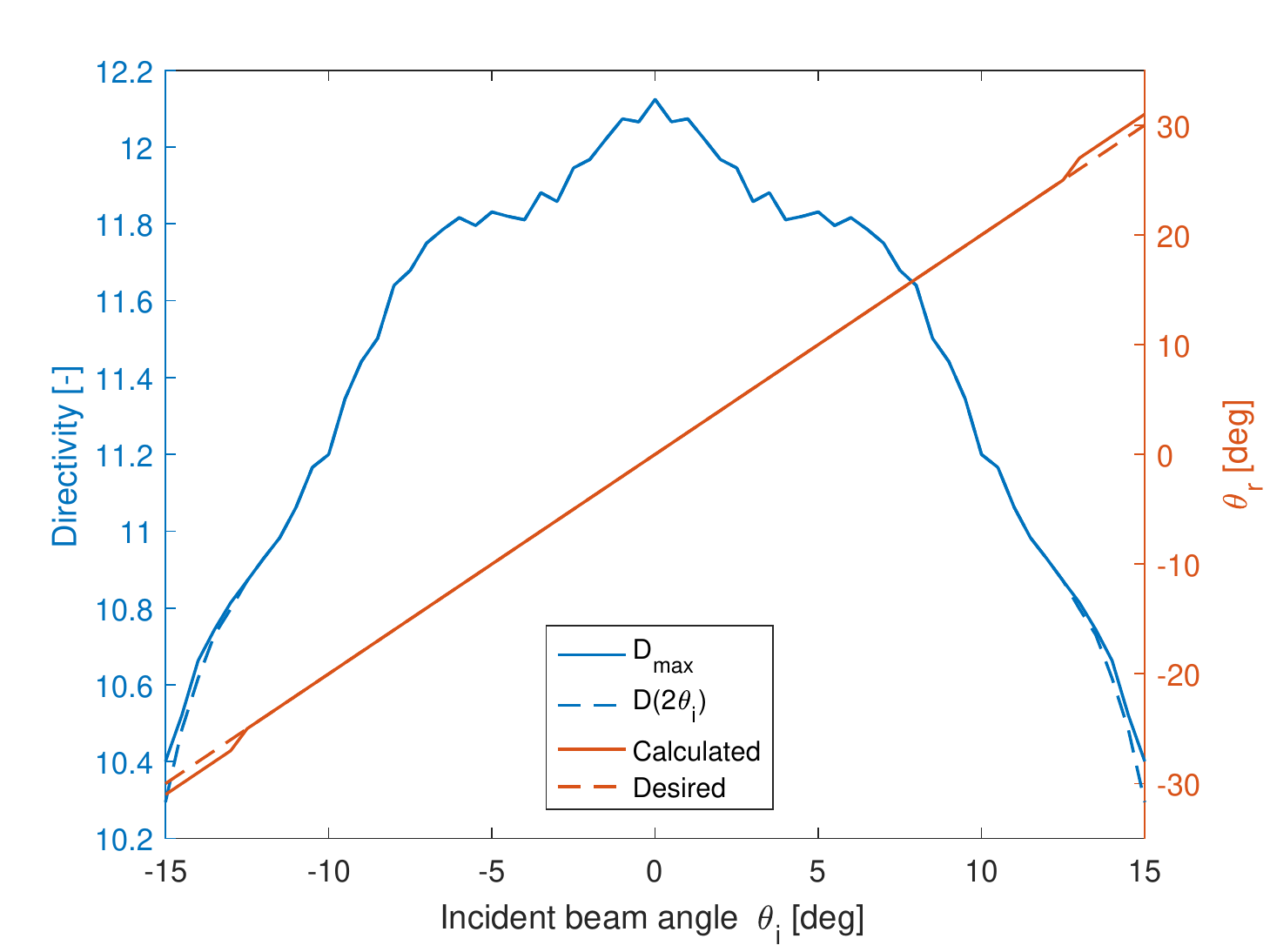}
  \caption{Scan and directivity performance of the angle-doubling phase boundary. The scan range of the array is extended by a factor of 2. Peak directivity and directivity at double the incidence angle are shown. Small errors in scan angle and similarity between the two directivity curves imply the phase boundary has a good performance.}\label{fig:PhaseSimDirPerfPhaseSurf}
\end{center}
\end{figure}
}

The ``$40\lambda$" curve of Fig. \ref{fig:AngleDoublerAlphavsD} compares the peak directivity from the surface with the peak directivity obtained from the array itself for a given angle of resulting beams. The ordinate axis is $\Delta D {=} D_{arr,\mathrm{dB}}(\theta){-}D_{surf,\mathrm{dB}}(\theta)$, where $D_{arr,\mathrm{dB}}(\theta)$ is the peak directivity of the array which is scanned to $\theta$ and $D_{surf,\mathrm{dB}}(\theta)$ corresponds to the values of the $D_\mathrm{max}$ curve of Fig. \ref{fig:PhaseSimDirPerfPhaseSurf}. Note that for the $D_{surf,\mathrm{dB}}(\theta)$ curve the array itself is scanned to $\theta/2$, while for $D_{arr,\mathrm{dB}}(\theta)$ the array is scanned to $\theta$. This was done to provide a meaningful comparison between the surface and the array itself in the full range of operation of the scan extending device. For an angle doubling setup with a distance of $40\lambda$ between the array and lens, at broadside the difference in directivity is 3.1dB. Theoretically the directivity difference should be 3dB for an angle-doubling lens. This is because the enhancement factor $\alpha$ is 2 and (\ref{eq:DirPerfAlpha}) reduces to $D'(2\theta)/D(\theta){=}1/2$, which corresponds to a 3dB reduction. The simulated results are close to this theoretical limit. The degradation in directivity of the surface stays roughly constant, as expected from the theory of Sec. \ref{sec:DirTheory}.

Distance of 40$\lambda$ between the source and the lens was considered in detail for future comparison -- because this is the largest distance which can be full-field simulated (see Sec. \ref{sec:MetaLens}) given our computational resources. Note however, that for the 16 element array, the far field (Fraunhofer) region starts at 112.5$\lambda$. Even though the $1/\alpha$ degradation of (\ref{eq:DirPerfAlpha}) was derived in the optical ray limit, the directivity of the beam peak still conforms to the expression even in the radiating near field of the source, as the ``$40\lambda$" curve of Fig. \ref{fig:AngleDoublerAlphavsD} depicts. In order to study how the $1/\alpha$ degradation of (\ref{eq:DirPerfAlpha}) for the peak of an incident beam behaves versus distance from the source, a number of angle doubling phase surfaces were simulated. Fig. \ref{fig:AngleDoublerAlphavsD} depicts the simulated data. Each curve corresponds to a different lens and a different array-lens distance, chosen to maintain the scan enhancement $\alpha$ of 2. The distance was varied from $10\lambda$ to $150\lambda$, which is well in the far field region. It appears that already at $30\lambda$ between the source and the lens, the direcitivity degradation is within 0.3dB of the expected 3dB at broadside. This suggests that at least for peaks of incident beams, the $1/\alpha$ directivity degradation is accurate well into the radiating near field of a source. Note that for the considered array-lens distances, apart from small errors, the angular scan performance of the doublers was not compromised.

\subsection{Simulating Angle-Tripling Phase Surfaces}

\begin{figure}[t!]
\begin{center}
\noindent
  \includegraphics[width=3.5in]{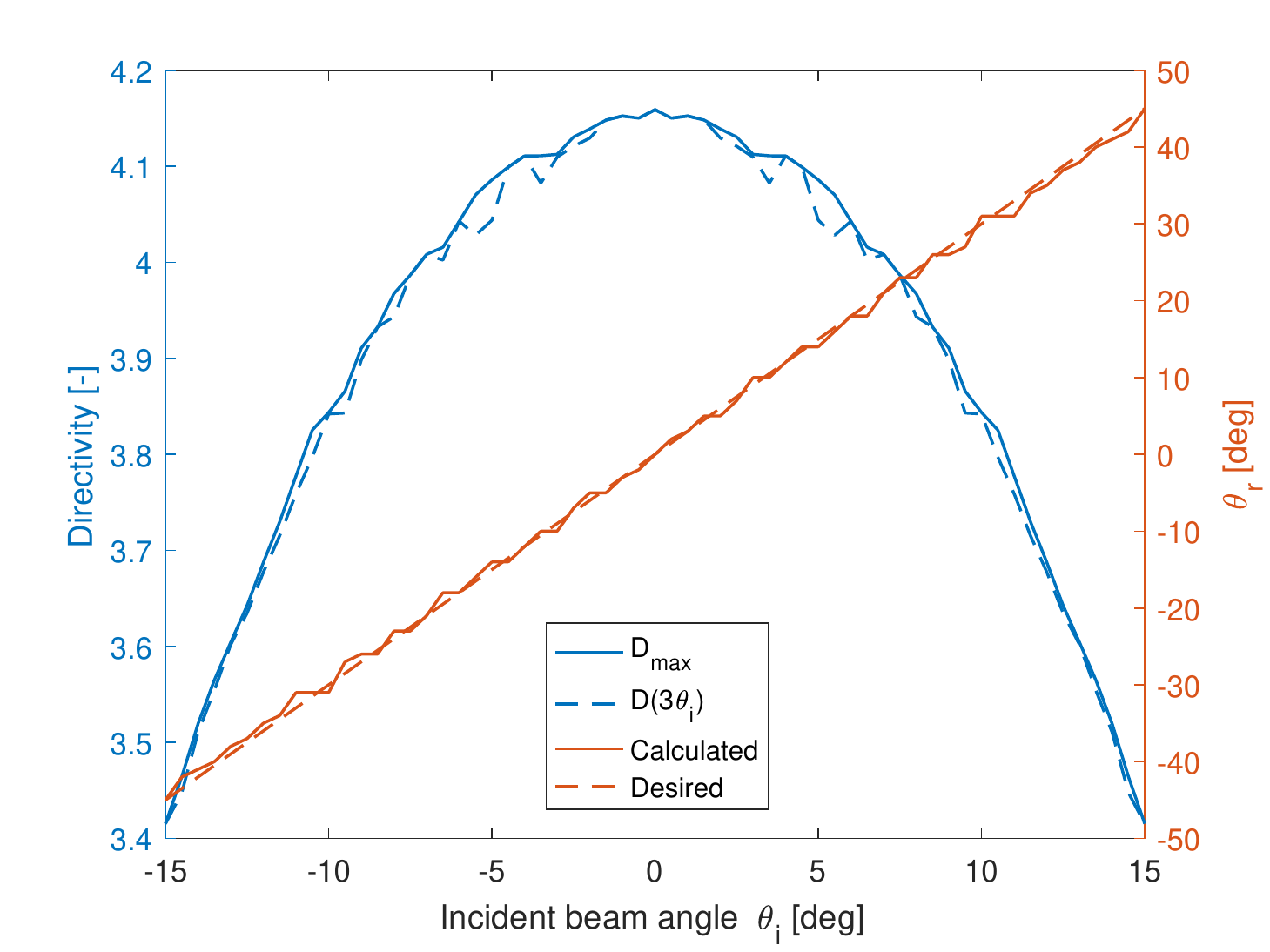}
  \caption{Scan and directivity performance of the angle-tripling phase boundary. The scan range of the array is extended by a factor of 3. Peak directivity and directivity at triple the incidence angle are shown. Small errors in scan angle and similarity between the two directivity curves imply the phase boundary has a good performance.}\label{fig:AngleTripDirPerf}
\end{center}
\end{figure}

For the sake of further verification of the presented theory, consider an angle-tripling lens ($\alpha{=}3$) excited by an 8 element, half-wavelength spaced antenna array. Initially, the distance between the array and the lens was chosen to be $30\lambda$, making the focal length of the lens equal $-15\lambda$. The array is still scanned to $\pm15^\circ$ off-broadside. Note that at $30\lambda$ the lens is in the far field (Fraunhofer) region of the 8 element source. Fig. \ref{fig:AngleTripDirPerf} depicts the scanning performance of the angle-tripling phase boundary. The error in output beam angles is within $\pm2^\circ$ compared to a perfect tripler. The discrepancy between peak directivity and the obtained directivity at triple of the incident angle is again insignificant.

\begin{figure}[t!]
\begin{center}
\noindent
  \includegraphics[width=3.5in]{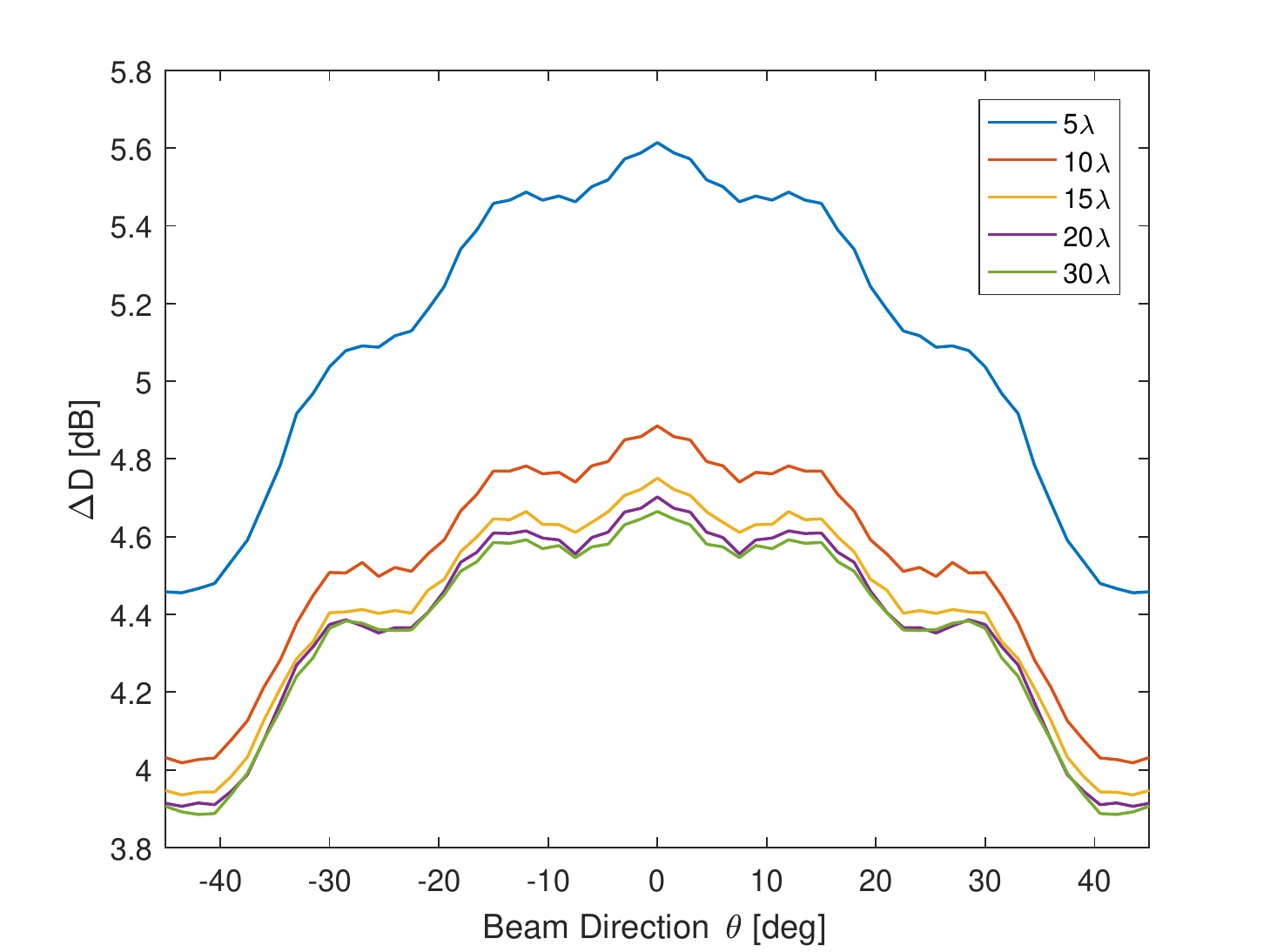}
  \caption{The directivity difference between the array itself and the scan extending lens at the same beam angle. Each curve corresponds to a different angle tripling scenario -- different lenses and array-lens separations. The legend gives the array-lens distance for each curve.}\label{fig:AngleTripDirComp}
\end{center}
\end{figure}

The ``$30\lambda$" curve of Fig. \ref{fig:AngleTripDirComp} compares the peak directivity from the $f{=}{-}15\lambda$ surface with the peak directivity obtained from the array itself for given angles of resulting beams. At broadside $\Delta D{=}4.7$dB, and stays within $4.6{\pm}0.1$dB for a large potion of the operating region. Note that this is in an approximate agreement with theory. According to (\ref{eq:DirPerfAlpha}), an angle-tripler is expected to exhibit a 4.8dB directivity degradation. The other curves of Fig. \ref{fig:AngleTripDirComp} correspond to directivity performance of different angle tripling scenarios, with different array-lens separations. It is again observed that good performance is obtained well into the radiating near field of the source.

\section{Design And Simulation of a Huygens' Omega Bianisotropic Angle-Doubling Lens} \label{sec:MetaLens}

The metasurface design methodology of \cite{epstein2016arbitrary} was used to design the angle doubler analagous to the one presented in Sec. \ref{sec:SimPhaseSurf}. Design and simulation were conducted at 10GHz. In order to completely specify the surface, the tangential electric and magnetic fields on both sides of the surface have to be postulated. The desired field transformation is then achieved via a sub-wavelength three-layer impedance structure, which emulates Omega bianisotropic scatterrers. Note that no physical unit cell design has been conducted \cite{epstein2016arbitrary}. As described in Sec. \ref{sec:LensEnh}, a diverging lens behaves as an angle doubler when the object is placed at its focal point. Thus the desired lens for this case study requires a focal length $f{=}{-}40\lambda$. It is well known that a diverging lens transforms a normally incident plane wave to a wave whose phase-fronts appear to originate at the focal point behind the lens. Thus we postulate the electric and magnetic fields on the transmission side of the lens ($y{=}0^+$) to be identical to the fields produced by an infinite line of current located at the focal point behind the lens. Using the geometry described above, the postulated transmitted electric and magnetic fields tangential to the surface are \cite{harrington1961time}
\begin{align}
E_z(x,0^+) &= \frac{k}{\omega\epsilon} \hphantom{,} \mathrm{H}_0^{(2)}\left(k\sqrt{x^2+f^2}\right), \\
H_x(x,0^+) &= \frac{\mathrm{j}f}{\sqrt{x^2+f^2}} \hphantom{,} \mathrm{H}_1^{(2)}\left(k\sqrt{x^2+f^2}\right),
\end{align}
where $\mathrm{H}_i^{(2)}$ is a Hankel function of second kind and order $i$. 

As discussed above, the fields on the incident side of the lens must resemble the fields of a normally incident plane wave. However, if one simply chooses the incident fields to be those of a normally incident uniform plane wave, real power flow across the surface will not be conserved \cite{epstein2016arbitrary}. This would lead to a lossy or active surface -- one which has to dissipate or supply power in order to facilitate the field transformation. In order to conserve real power flow across the surface, the incident fields are postulated as
\begin{align}
E_z(x,0^-) &= \sqrt{\eta \Re{E_z(x,0^+)H_x^*(x,0^+)}},\\
H_x(x,0^-) &= \frac{E_z(x,0^-)}{\eta}.
\end{align}
This choice of incident fields makes sure that the phase of electric field along the surface is constant and the real part of the normal component of the Poynting vector is equal on both sides of the surface, i.e. $\Re{S_y(x,0^+)}=\Re{S_y(x,0^-)}$. It is worth mentioning that the postulated field transformation described in this section leads to a phase shift which is identical to the one provided in (\ref{eq:LensPhase}). Also note that the postulated field transformation presented in this section does not maintain the magnitudes of $E_z(x,0^\pm)$ and $H_x(x,0^\pm)$ across the metasurface, and thus does not conform to the previously discussed phase boundary transformation of Sec. \ref{sec:PhaseBoundary}. Here a physical bianisotropic lens is designed, while (\ref{eq:PhaseBoundaryE}) and (\ref{eq:PhaseBoundaryH}) were introduced for the benefit of simpler and faster simulations. Thus, it is not necessary for a physical surface design to conform to  the phase boundary transformation. 

\begin{figure}[t!]
\begin{center}
\noindent
  \subfigure[]{\includegraphics[width=3.5in]{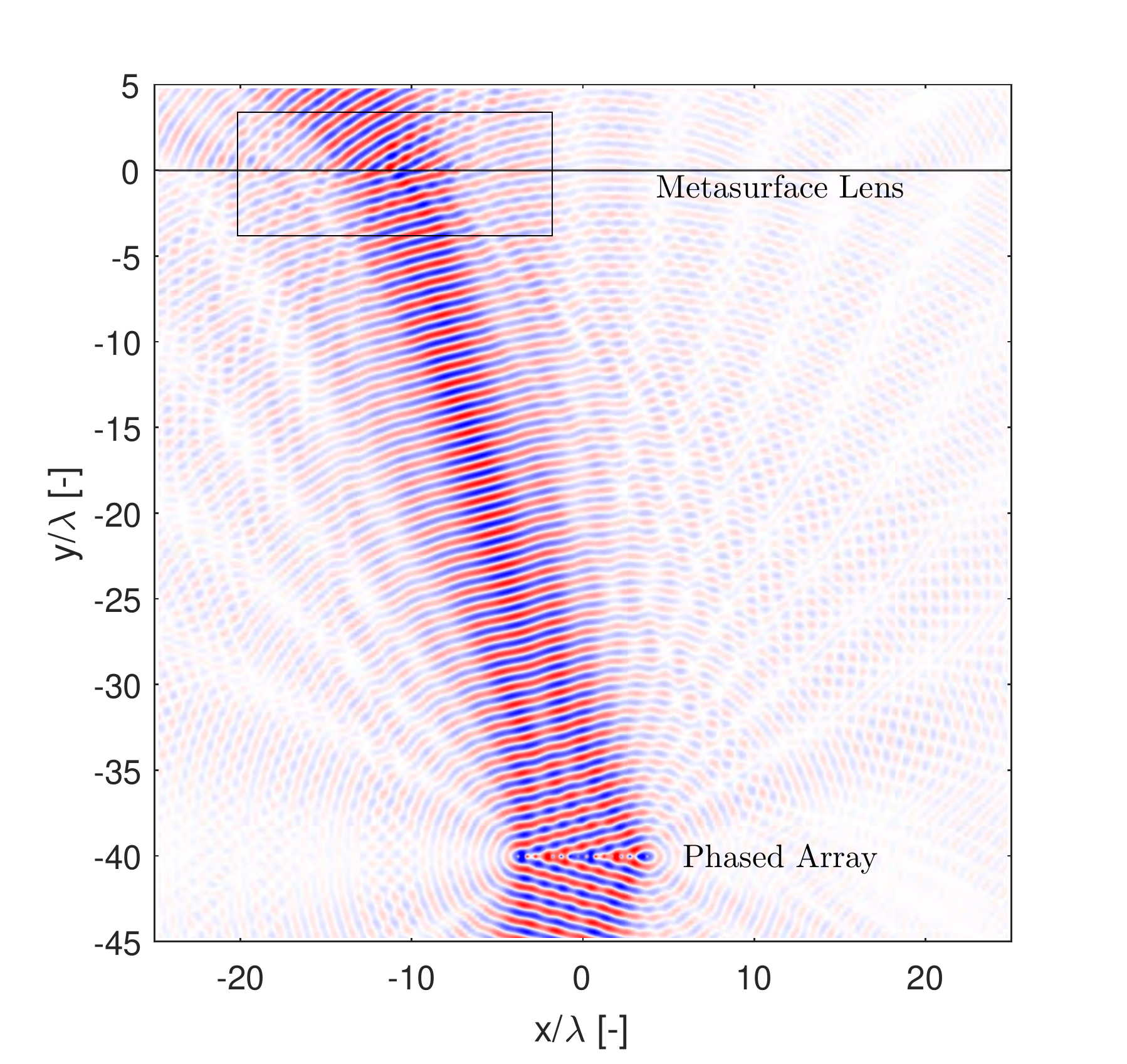}}
  \subfigure[]{\includegraphics[width=3.5in]{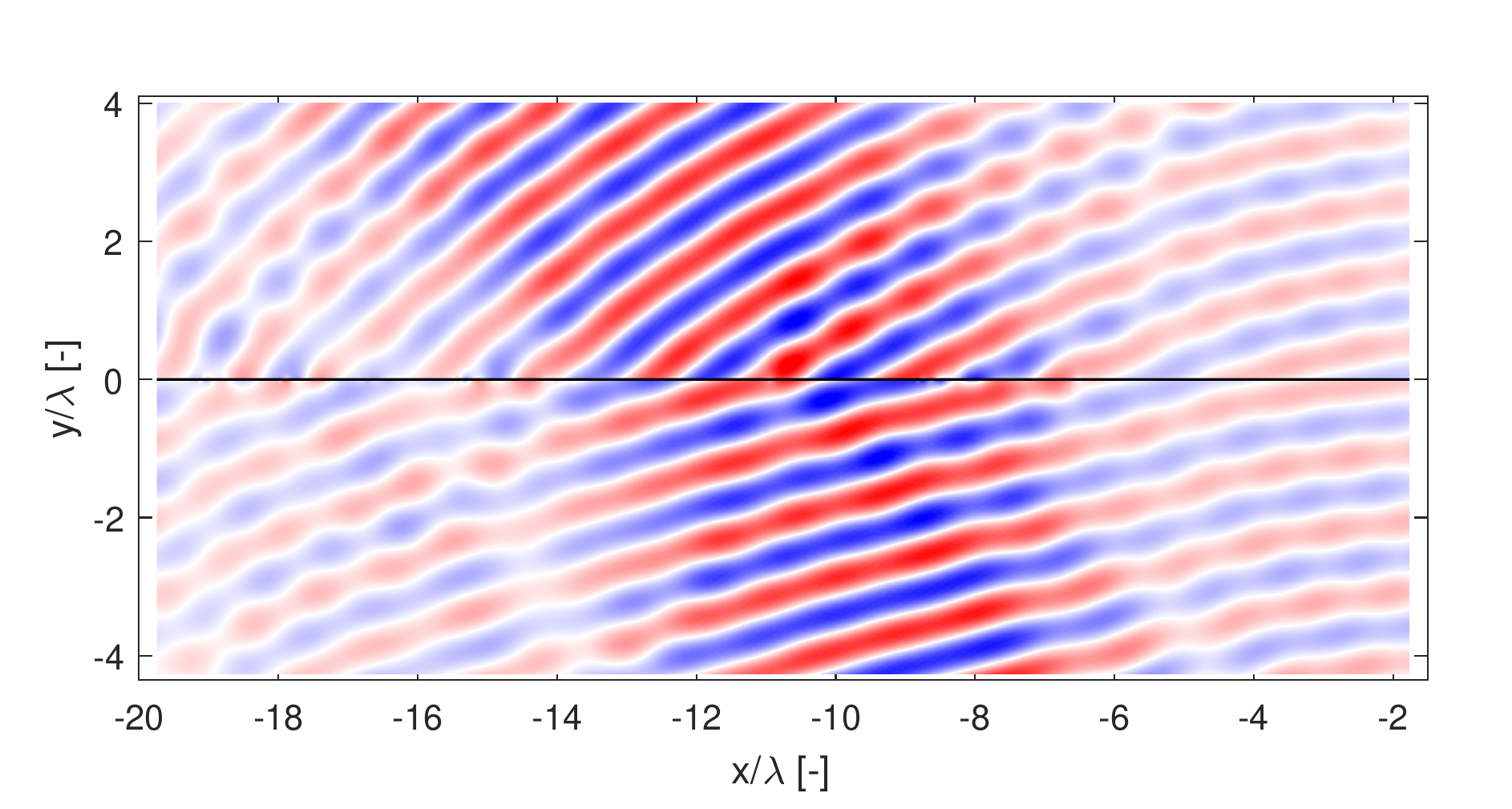}}
  \caption{Simulated behavior of the proposed angle-doubling metasurface lens. Simulated via COMSOL Multiphysics. The plots depict the out-of-plane electric field. (a) Depicts the complete simulation domain. The 16 element phased array source is labeled, as well as the metasurface lens. The lens and array are separated by 40$\lambda$. The array produces a 15$^\circ$ off-broadside beam, which is refracted to 30$^\circ$ off-broadside. It is clearly visible that the lens performs in a nearly lossless and reflectionless manner. The rectangle placed around the beam refraction location shows the focus of (b). 
  }\label{fig:LensArrayFields}
\end{center}
\end{figure}

The three-layer impedance structure which facilitates the postulated field transformation was programmed and simulated in COMSOL Multiphysics. Fig. \ref{fig:LensArrayFields} depicts the scattering behavior of the designed lens when illuminated by a beam from the phased array. It must be noted that although the surface is illuminated with a field that is quite different from the postulated one, the surface nevertheless appears to perform well as a lens without significant reflections or losses. Furthermore, the favorable performance is maintained as the phased array scans its beam. It is remarkable how well the metasurface lens performs when illuminated by these varying fields. So far Huygens' metasurfaces were designed and shown to perform well for a single incidence scenario only. Here not only is the surface not illuminated by the postulated incident field, the incident field itself varies with angle and yet the surface performs as expected from a diverging lens. 

\begin{figure}[t!]
\begin{center}
\noindent
  \includegraphics[width=3.5in]{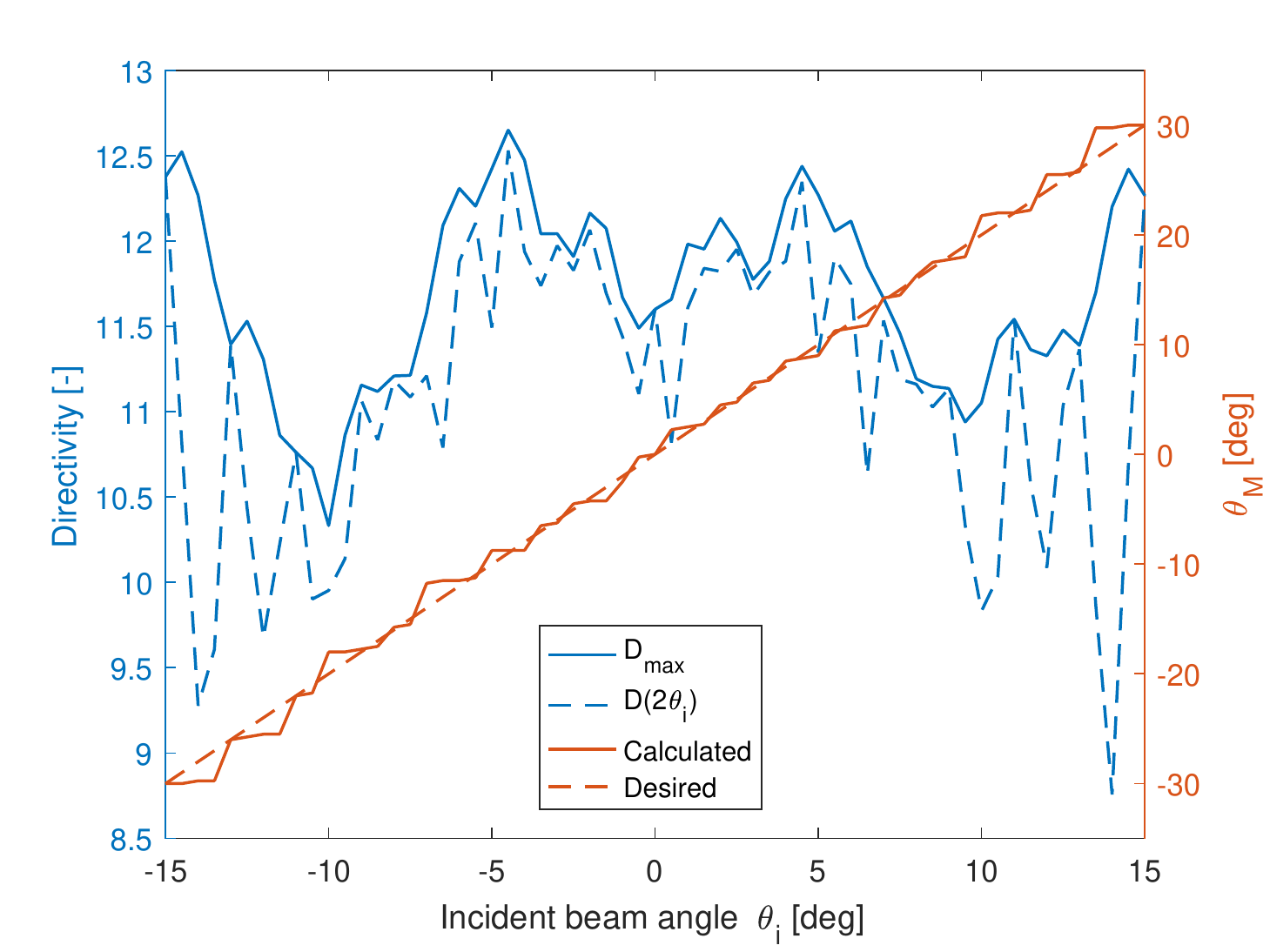}
  \caption{Replicated result of Fig. \ref{fig:PhaseSimDirPerfPhaseSurf} with COMSOL full-wave simulation of the bianisotropic metasurface.}\label{fig:DirPerf}
\end{center}
\end{figure}

\begin{figure}[t!]
\begin{center}
\noindent
  \includegraphics[width=3.5in]{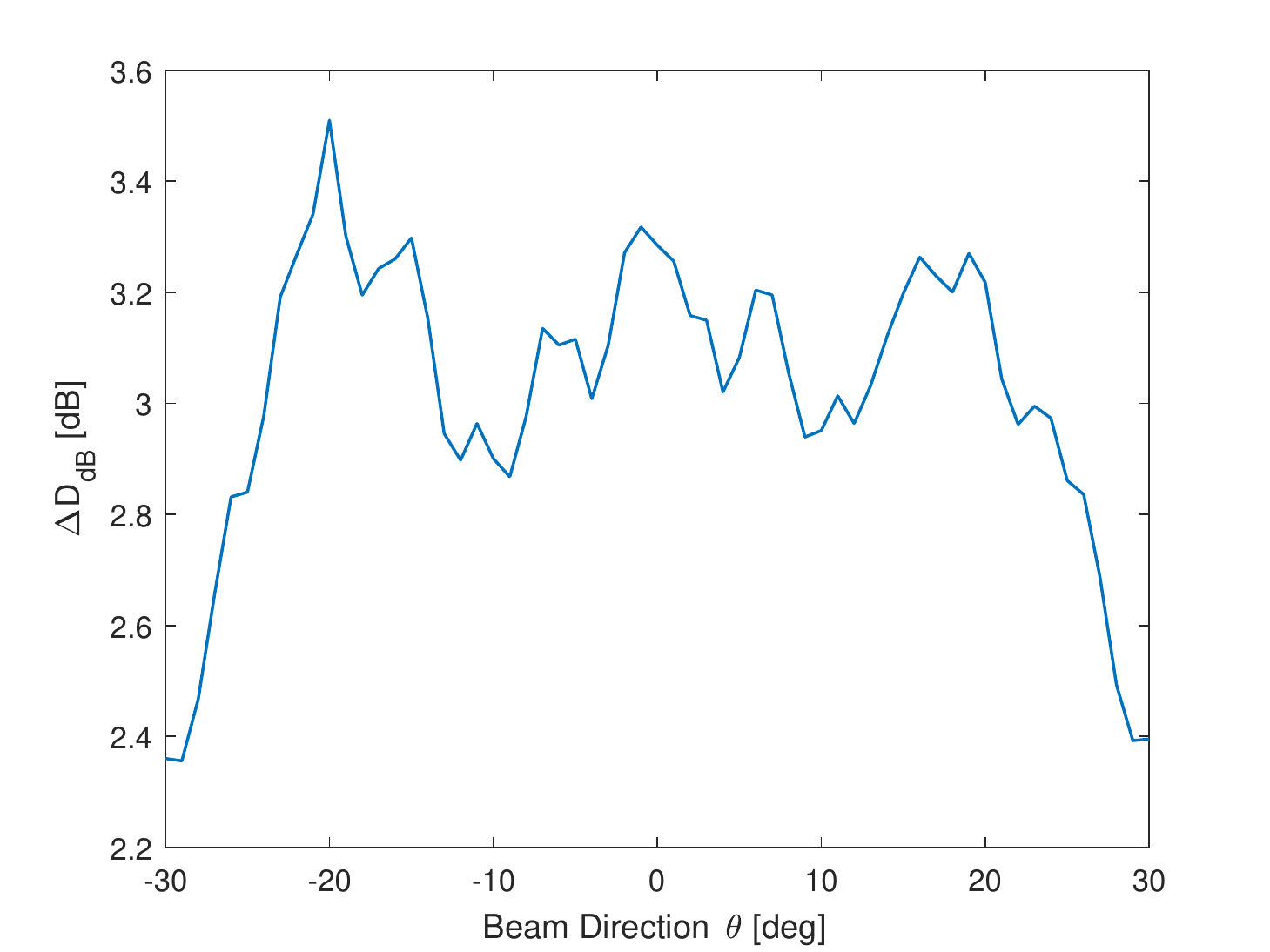}
  \caption{Replicated result of the ``$40\lambda$" curve of Fig. \ref{fig:AngleDoublerAlphavsD} with COMSOL full-wave simulation of the bianisotropic metasurface.}\label{fig:DirComp}
\end{center}
\end{figure}

Figs. \ref{fig:DirPerf} and \ref{fig:DirComp} reproduce the results of Fig. \ref{fig:PhaseSimDirPerfPhaseSurf} and the ``$40\lambda$" curve of Fig. \ref{fig:AngleDoublerAlphavsD}, but using far-field data obtained with the full-wave simulation of the designed Omega bianisotropic metasurface. It is clear that the results of the two simulation methods are in close agreement. This observation validates the described simulation method of Sec. \ref{sec:PhaseBoundary} and the theory of Sec. \ref{sec:DirTheory}.

\section{Conclusion}

This paper has considered the problem of extending the scan capabilities of antenna arrays via a metasurface lens. It was first shown that a diverging lens can achieve an arbitrary array scan enhancement but at the cost of reduced directivity. This behavior was quantified by defining directivity in the context of ray optics and studying how a scan extending lens affects it. An analytic relationship between the angular scan enhancement and directivity behavior was derived. It was also shown for the specific case of broadside incidence upon an angle-doubling lens that it is possible to place the lens in the near-field of the illuminating array while still suffering the same directivity degradation as for the far-field case, albeit with some limitations. 

A common simulation approach for scan enhancing metasurfaces, which treats the surfaces as phase boundaries, was discussed in detail. For this, an interpretation of a phase boundary in the context of electromagnetics was also presented. To verify the theoretical claims of the paper, angle-doubling and angle -tripling lenses were simulated with this phase boundary method, which showed close agreement between theory and simulated behavior.

To verify the results of the phase boundary simulations and to depict the physical possibility of a low-loss near-reflectionless metasurface lens, an angle-doubling metasurface lens was designed by postulating tangential fields on both sides of the surface. The resulting Omega bianisotropic Huygens' metasurface design was analyzed via COMSOL Multiphysics simulations. The simulation results showed how the metasurface functions as a diveriging lens for all incident fields of interest. Throughout the operating scan range of the exciting phased array, the surface remained low-loss and near-reflectionless. The surface performance was in close agreement with the simulation results of Sec. \ref{sec:SimPhaseSurf}.

\appendices

\section{Directivity in 2D Ray Optics} \label{sec:AppRayD2D}

In two dimensions, fields due to sources decay as $1/\sqrt{r}$ ($r$ being the distance from the source to an observation point), whereas in three dimensions the decay has $1/r$ dependence \cite{harrington1961time}. Furthermore, in two dimensions all possible directions from a source are spanned by a single angle $\theta$, while directions in three dimensions require two independent angles. Because of these differences, the two-dimensional radiation intensity is defined as
\begin{equation}
U_{2D}(\theta){=}\lim_{r\to\infty} r \frac{|\mathbf{E}(r,\theta)|^2}{2\eta},
\end{equation}
where $\mathbf{E}(r,\theta)$ is the electric field phasor vector produced by the sources. The average (with respect to $\theta$) radiated power can be written as $\int_0^{2\pi} U_{2D}(\theta)\d\theta/{2\pi}$, with units of W/m. An analog of three-dimensional directivity in two dimensions can now be written as 
\begin{equation} \label{eq:AppDirInf}
D_{r\to\infty}(\theta)=\frac{2\pi U_{2D}(\theta)}{\int U_{2D}(\beta) \d\beta}.
\end{equation}
The $r{\to}\infty$ subscript signifies that the required power densities are calculated in the limit of infinite $r$.

We are interested in how a lens, which has definite position along the optical axis from a source, affects the radiation characteristics of said source. Thus the directivity in terms of quantities which are evaluated infinitely far away is not adequate for analyzing the effect of the lens. This issue does not appear in the context of ray optics. This occurs because ray optics itself is obtained from electromagnetics via the limit of $\lambda{\to}0$ \cite{landau2013classical}. To see why this is the case, consider the radiation from a source located near the origin. For many wavelengths away from the origin ($r {\gg} \lambda$) the radiation can be written in the form \cite{harrington1961time,vladimirov1976equations}
\begin{equation} \label{eq:AppEmagSource}
|\mathbf{E}(r,\theta)| = \frac{f(\theta)}{\sqrt{r}}.
\end{equation}
By default, in the ray optical limit ($\lambda{\to}0$) any finite distance from the source is infinite in terms of the number of wavelengths, and therefore (\ref{eq:AppEmagSource}) is valid at any finite distance away from the source. The ray density, which is a ray optical concept, can now be related to the two-dimensional radiation intensity by setting $\rho(\theta){=}f^2(\theta)$. With this choice of ray density, it is easy to see that $U_{2D}(\theta) \propto \rho(\theta)$, which makes the directivities of (\ref{eq:2DDirRho}) and (\ref{eq:AppDirInf}) identical to each other. In ray optics one can trace an arbitrary number of rays from a source with a given $\rho(\theta)$. These rays have local character and changes to $\rho(\theta)$ can be analyzed in a local sense as was done in Sec. \ref{sec:DirTheory}.

\section{Peak Directivity of a 2D Aperture} \label{sec:AppDUmax}
A well known result from basic antenna theory is the peak directivity of a uniform aperture of area $A$, which is \cite{balanis2016antenna}
\begin{equation}
D_{3D,\mathrm{max}} = \frac{4\pi A}{\lambda^2}.
\end{equation}
The two-dimensional analog of the above equation is obtained as follows. The ray density of a two-dimensional uniform aperture of length $L$ is
\begin{equation}
\rho_U(\theta) \propto
\begin{cases}
	\sinc^2 \left(\frac{kL}{2}\sin\theta \right) \cos\theta,& \text{if } |\theta| \leq \frac{\pi}{2}\\
	0, & \text{otherwise}.
\end{cases}
\end{equation}
This is a well-known result, which is obtained via the magnitude squared of a one-dimensional Fourier transform of the fields over the two-dimensional rectangular aperture \cite{landau2013classical}.
Using the above ray density in (\ref{eq:2DDirRho}),
\begin{equation} \label{eq:AppDU}
D_U(\theta)=\frac{2\pi \rho_U(\theta)}{\int_{-\pi/2}^{\pi/2} \sinc^2 \left(\frac{kL}{2}\sin\beta \right) \cos \beta \hphantom{,} \d\beta}. 
\end{equation}
Assuming $L {\gg} \lambda$, the integral appearing in the above equation can be approximated as
\begin{multline}
\int_{-\pi/2}^{\pi/2} \sinc^2 \left(\frac{kL}{2}\sin\beta \right) \cos\beta \hphantom{,} \d\beta \approx \\
\approx \int_{-\infty}^\infty \sinc^2 \left( \frac{kL}{2}\beta \right) \d\beta=\frac{\lambda}{L}.
\end{multline}
The ray density peaks at $\theta{=}0$, thus $D_{U,\mathrm{max}}{=}D_U(0)$. Using the above integral approximation,
\begin{equation} \label{eq:AppDUmax}
D_{U,\mathrm{max}}=\frac{2\pi L}{\lambda}.
\end{equation}
It is worth mentioning how well this approximation holds even when the aperture length is on the order of a wavelength. Equation (\ref{eq:AppDU}) was numerically evaluated for the case $\theta{=}0$ and $L{=}2\lambda$, and the result was compared with the value obtained via (\ref{eq:AppDUmax}). The discrepancy between the two values is 5\%.

\section{Phase Boundary Definition: a Closer Look} \label{sec:AppPhaseCloserLook}

Consider the scattering of ideal plane waves by a linear-phase surface whose phase function is $\phi(x){=}k_s x$. The fields of an incident TE plane wave are given by \cite{harrington1961time}
\begin{align}
\mathbf{E}_i(x,y{<}0) &= \hat{\mathbf{z}}E_i e^{-\mathrm{j}(k_{i,x}x+k_{i,y}y)}, \label{eq:EincPW} \\ 
\mathbf{H}_i(x,y{<}0) &= \frac{E_i}{\eta k} e^{-\mathrm{j}(k_{i,x}x+k_{i,y}y)}
\begin{bmatrix}
k_{i,y} \\
-k_{i,x} \\
0
\end{bmatrix} \label{eq:HincPW}
\end{align}
where $k_{i,x}$ and $k_{i,y}$ are the $x$- and $y$-components of the incident wave vector. The phase surface transforms the tangential components of these incident fields into the transmitted tangential fields, given by:
\begin{align}
E_z(x,0^+) &= E_i e^{-\mathrm{j}(k_{i,x}-k_s)x}, \label{eq:EtransPW} \\ 
H_x(x,0^+) &= \frac{E_i k_{i,y}}{\eta k} e^{-\mathrm{j}(k_{i,x}-k_s)x}. \label{eq:HtransPW}
\end{align}

As was mentioned above, there exists an Omega bianisotropic metasurface which achieves this field transformation. Now, due to the concrete nature of incident fields and surface phase, we are in a position to discuss exactly what this field transformation achieves in regions away from the surface itself. Fields away from the surface must satisfy Maxwell's equations. Thus, we are interested in the exact form of the fields in the $y{<}0$ and $y{>}0$ regions which satisfy Maxwell's equations and the boundary tangential fields along the surface. For the $y{<}0$ region the result is obvious and is given by (\ref{eq:EincPW}) and (\ref{eq:HincPW}). For the $y{>}0$ region, it is found that a combination of two planes waves is required. The fields of these waves have the form
\begin{align}
\mathbf{E}_{1,2}(x,y{>}0) &= \hat{\mathbf{z}}E_{1,2} e^{-\mathrm{j}\left((k_{i,x}-k_s)x \pm k_{y}y\right)}, \label{eq:Ewaves12} \\ 
\mathbf{H}_{1,2}(x,y{>}0) &= \frac{E_{1,2}}{\eta k} e^{-\mathrm{j}\left((k_{i,x}-k_s)x \pm k_{y}y\right)}
\begin{bmatrix}
\pm k_y \\
k_s -k_{i,x} \\
0
\end{bmatrix}, \label{eq:Hwaves12}
\end{align}
where $k_{y}{=}\sqrt{k^2-(k_{i,x}-k_s)^2}$. Subscripts 1 and 2 correspond to the two plane waves, and the plus sign is taken for wave 1 and minus for wave 2. The amplitudes of these two waves have to equal
\begin{equation} \label{eq:E12ampl}
E_{1,2} = \frac{1}{2}\left(1\pm\frac{k_{i,y}}{k_{y}}\right)E_i.
\end{equation}

These fields correspond to a wave with amplitude $E_1$ traveling away from the surface and another one with amplitude $E_2$ traveling towards the surface. This behavior depicts another inconsistency of the defined phase boundary. The surface is meant to be excited by a single plane wave from the $y{<}0$ region. However, the desired phase discontinuity is achieved in a rigorous fashion with a physical Omega bianistropic metasurface only with two incident plane waves, the other being incident from the $y{>}0$ region. Note that for small values of $|k_s|$ and as long as $k_{y}$ remains real, $k_{i,y}/k_{y} {\approx} 1$, which makes $E_1 {\approx} E_i$ and $E_2 {\approx} 0$.

The phase boundary simulation procedure described in Sec. \ref{sec:SimPhaseSurf} does not have any sources outside of the equivalence box. Thus, phase boundary simulations are incapable of producing wave 2 of (\ref{eq:Ewaves12}) and (\ref{eq:Hwaves12}), making the obtained fields in the $y{>}0$ region unphysical. This leads to the question of what actually happens when a phase boundary is simulated? We answer this question by studying the radiated fields of the equivalent currents at the linear phase boundary. The simulation method uses the transmitted fields of (\ref{eq:EtransPW}) and (\ref{eq:HtransPW}) to obtain Love's equivalent surface currents at the surface, which are
\begin{align}
\mathbf{J}(x,0) &= -\hat{\mathbf{z}}\frac{E_i k_{i,y}}{\eta k} e^{-\mathrm{j}(k_{i,x}-k_s)x}, \label{eq:Jequiv}\\
\mathbf{M}(x,0) &= -\hat{\mathbf{x}}E_i e^{-\mathrm{j}(k_{i,x}-k_s)x}. \label{eq:Mequiv}
\end{align}
Using symmetry arguments one can deduce that the electric current $\mathbf{J}(x,0)$ on its own produces two planes waves, one in the $y{<}0$ region and the other in $y{>}0$ region, both of which propagate away from the $y{=}0$ surface. The wave vectors of the two plane waves are $\mathbf{k}_\lessgtr{=}\hat{\mathbf{x}}(k_{i,x}-k_s) \mp \hat{\mathbf{y}}k_y$, where $\mathbf{k}_\lessgtr$ is for the plane wave in the $y \lessgtr 0$ region and the minus sign is used for $\mathbf{k}_<$. The same can be said of the magnetic current $\mathbf{M}(x,0)$. Acting together, the two solutions superimpose to produce a single plane wave in the $y{<}0$ region and a single plane wave in the $y{>}0$ region, with the fields $\mathbf{E}_\lessgtr(x,y \lessgtr 0)$, $\mathbf{H}_\lessgtr(x,y \lessgtr 0)$. 

These two plane waves must satisfy boundary conditions imposed by the currents, which become
\begin{align}
\hat{\mathbf{y}}\times \left( \mathbf{H}_>(x,0^+)-\mathbf{H}_<(x,0^-) \right) &= \mathbf{J}(x,0), \label{eq:Jboundary}\\
\left( \mathbf{E}_>(x,0^+)-\mathbf{E}_<(x,0^-) \right) \times \hat{\mathbf{y}} &= \mathbf{M}(x,0). \label{eq:Mboundary}
\end{align}
One can easily check that choosing 
\begin{align}
\mathbf{E}_> &= \mathbf{E}_1,& \mathbf{H}_> &=\mathbf{H}_1,\\ 
\mathbf{E}_< &= -\mathbf{E}_2,& \mathbf{H}_< &= -\mathbf{H}_2
\end{align}
satisfies the boundary conditions. Note that the coordinate dependence of the fields was omitted for brevity and $\mathbf{E}_2(x,y{<}0)$, $\mathbf{H}_2(x,y{<}0)$ are given by the same expressions as $\mathbf{E}_2(x,y{>}0)$, $\mathbf{H}_2(x,y{>}0)$.

We summarize these results as follows. When the equivalence principle was applied it was assumed that no sources are present beyond the box surrounding the array and metasurface. However, a closer inspection of the tangential transmitted fields has shown this assumption to be incorrect. Because of this, the equivalent surface currents on the transmission side of the lens radiate into both $y{>}0$ and $y{<}0$ regions. For $y{>}0$, these currents produce wave 1 of (\ref{eq:Ewaves12}) and (\ref{eq:Hwaves12}). For $y{<}0$, a wave is produced which has the same form as wave 2 of (\ref{eq:Ewaves12}) and (\ref{eq:Hwaves12}), but whose amplitude is the negative of $E_2$. 

\begin{figure}[t!]
\begin{center}
\noindent
  \includegraphics[width=3.5in]{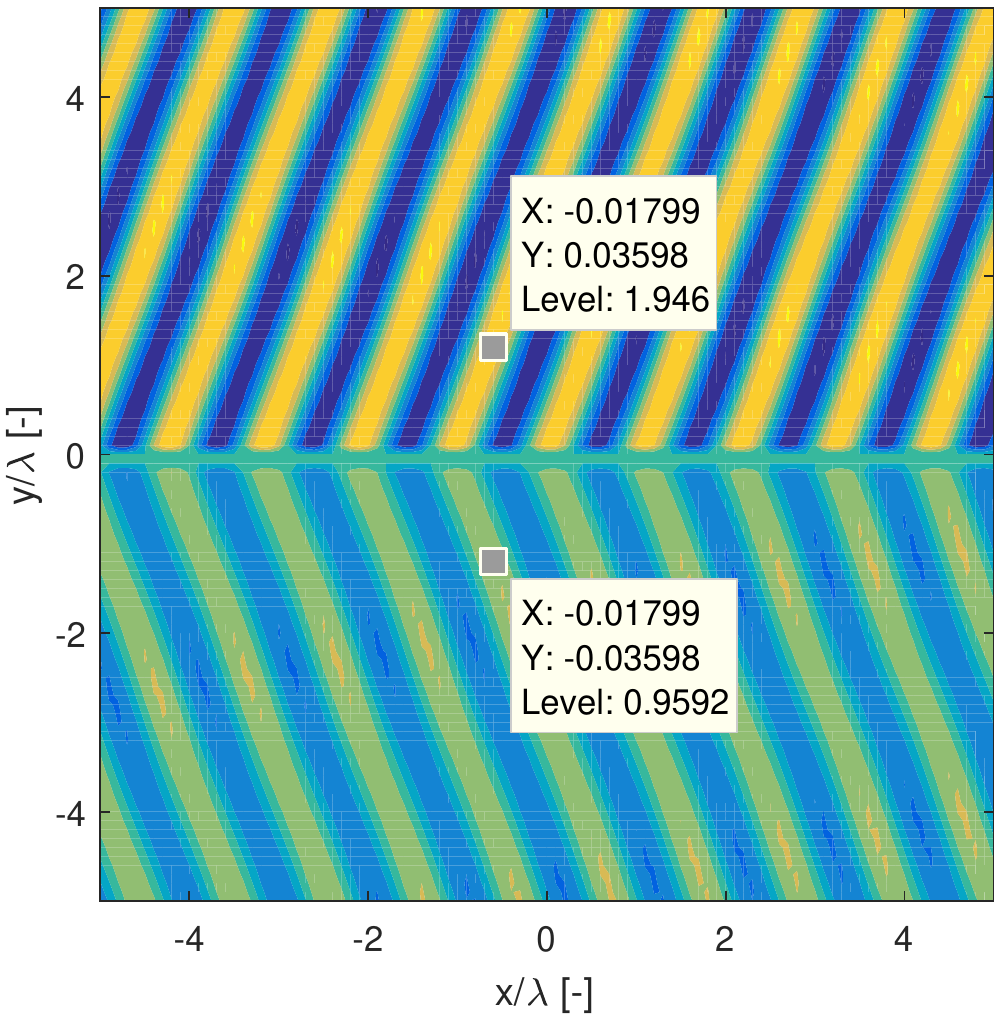}
  \caption{Radiated fields by equivalent currents which result due to a normally incident plane wave onto the phase boundary, which is described by $\phi(x){=}k \sin(70^\circ) x$. The equivalent currents are located at $y{=}0$ and limited to $|x|{\leq}200\lambda$. The equivalence box is not labelled, and the fields are not plotted inside it.} \label{fig:JMequivRad}
\end{center}
\end{figure}

Let us now show that this indeed occurs in phase boundary simulations. For this we consider a scenario where the equivalence box is a rectangle with its four sides along the $y{=}0$, $y{=}{-}0.2\lambda$, $x{=}{\pm}200\lambda$ planes. We assume that within the equivalence box there are sources which produce incident fields on the phase boundary to be those of a normally incident plane wave. The phase boundary is given by $\phi(x){=}k \sin(70^\circ) x$. This boundary is meant to refract the normally incident plane wave towards 70$^\circ$ off-broadside. Non-zero equivalent currents appear only on the side of the equivalence box which touches the phase boundary. Thus, the simulated fields are only due to the equivalent currents on the $y{=}0$ plane (limited by $|x|{\leq} 200\lambda$). According to (\ref{eq:E12ampl}) we expect $E_i{=}1$, $E_1{=}1.96$, and ${-}E_2{=}0.96$. Figure \ref{fig:JMequivRad} shows the fields obtained by such a simulation. Indeed the equivalent currents produce two plane waves in the $y \lessgtr 0$ regions which exhibit the expected amplitudes.

The presented analysis of what happens when a linear phase boundary is simulated under plane wave incidence leads to three more inconsistencies regarding the phase boundary definition of (\ref{eq:PhaseBoundaryE}) and (\ref{eq:PhaseBoundaryH}). First of all, the phase boundary definition is meant to produce tangential fields at $y{=}0^+$ given by (\ref{eq:EtransPW}) and (\ref{eq:HtransPW}). This however is not the case, since the actual fields at $y{=}0^+$ produced by the phase boundary simulations are $\mathbf{E}_1(x,0^+)$ and $\mathbf{H}_1(x,0^+)$. The second inconsistency is the fact the surface is meant to behave in a reflectionless manner. In reality, phase boundary simulations exhibit reflections as demonstrated by the existence of $\mathbf{E}_<(x,y{<}0)$ and $\mathbf{H}_<(x,y{<}0)$. The third inconsistency stems from the fact that the phase boundary definition was chosen such that a physical Omega bianisotropic metasurface corresponding to a desired phase boundary would be passive and lossless. The above expressions show that in simulation the linear phase boundary can be active or lossy. This can be observed by considering the incident, transmitted and reflected power from a section of the phase boundary of length 1:
\begin{align}
P_i &= \frac{E_i^2 k_{i,y}}{2\eta k}, &
P_t &= \frac{E_1^2 k_y}{2 \eta k}, &
P_r &= \frac{E_2^2 k_y}{2 \eta k}.
\end{align}
Plugging in the expressions for $E_1$ and $E_2$ into the above we obtain
\begin{equation}
P_t+P_r = \frac{k_y}{4\eta k} \left( 1 + \frac{k_{i,y}^2}{k_y^2} \right) E_i^2.
\end{equation}
For the case of $k_s{\neq}0$, $P_i \neq P_t + P_r$, which implies the surface is either active or lossy.

Now consider a scenario where $\hat{\mathbf{k}}_i$ makes an angle of $\theta$ with $\hat{\mathbf{y}}$ and the surface is chosen in such a way as to enhance this angle to $\theta'{>}\theta$. For this case (\ref{eq:E12ampl}) becomes
\begin{equation}
E_{1,2} = \frac{1}{2}\left(1 \pm \frac{\cos\theta}{\cos\theta'}\right)E_i.
\end{equation}
For $\theta{=}15^\circ$ and $\theta'{=}30^\circ$, which approximates the most extreme incidence case of the angle-doubling lens simulations, $E_1{=}1.06 E_i$ and $E_2{=}{-}0.06 E_i$. The small magnitude of $E_2$ in this case explains why the results of Sec. \ref{sec:MetaLens} and Sec. \ref{sec:SimPhaseSurf} agree. Even though the simulation of the phase boundary lens produces spurious reflections and artificially enhances the amplitude of the refracted beam, these effects are small. The spurious reflection with the amplitude of approx. $-E_2$ appears in the region $y{<}0$ outside of the bounding box, but its amplitude is too weak to be noticed in Fig. \ref{fig:PhaseSim15to30}.

\begin{figure}[t!]
\begin{center}
\noindent
  \subfigure[]{\includegraphics[width=3.5in]{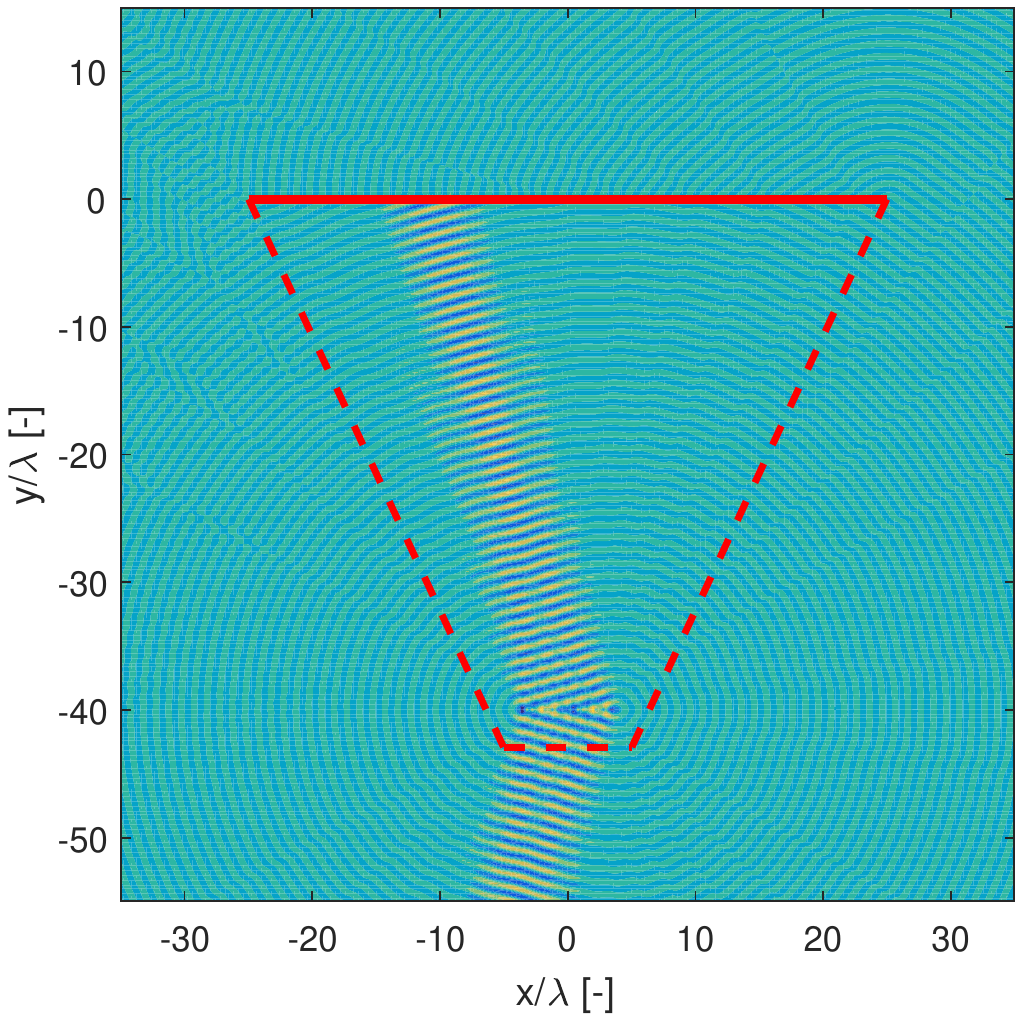}}
  \subfigure[]{\includegraphics[width=3.5in]{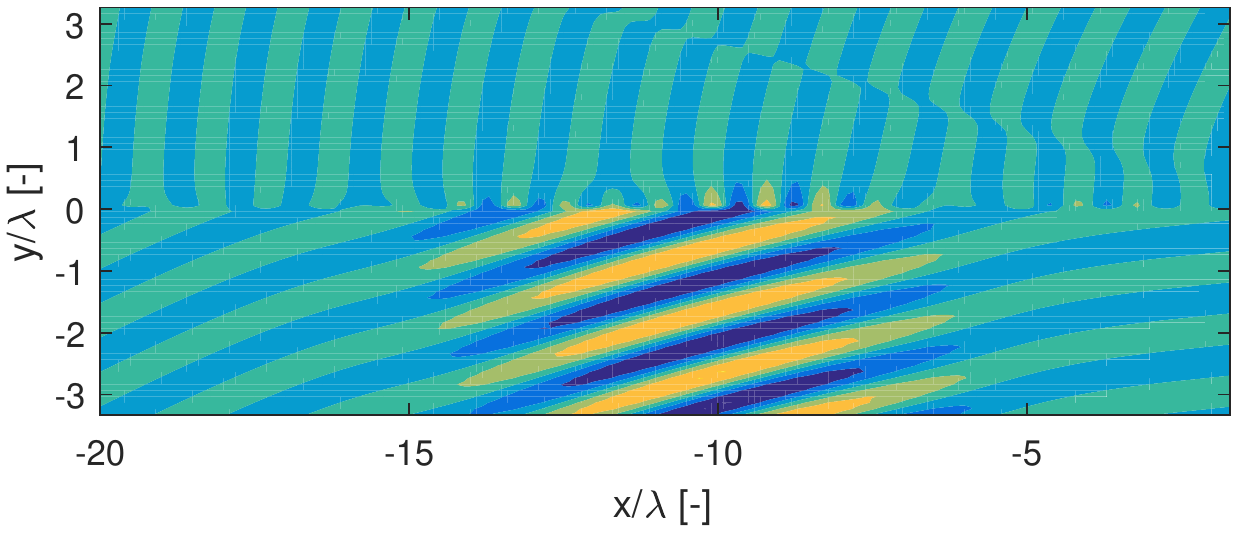}}
  \caption{A 15$^\circ$ off broadside beam illuminating a $\phi(x){=}0.95kx$ phase surface. (a) clearly depicts the loss of radiating array power as the transmitted fields become evanescent. (b) zooms in on the radiating beam to evanescent fields transition at the phase surface.
  }\label{fig:PhaseSim15toEvanesc}
\end{center}
\end{figure}

Perhaps the most interesting artifact of the electromagnetic phase boundary simulations is the possible disappearance of radiating array power. Up to now our discussion of the fields in the $y{>}0$ region was limited to the case where $|k_{i,x}-k_s|{<}k$. The obtained expressions of the two plane waves remain valid even if this condition is relaxed. In this case $k_{y}$ can be written as $-\mathrm{j}\alpha$, where $\alpha{\geq}0$. This results in two evanescent waves, one attenuating and another growing away from the surface. The growing evanescent wave further reinforces the fact that sources must be present in the $y{>}0$ region to achieve the desired phase transformation. In simulation, the growing evanescent wave appears as a decaying evanescent wave in the $y{<}0$ region. This effect becomes possible when the incident beam/plane wave angle with respect to the broadside direction is large and/or when the surface exhibits strong refraction ($k_s$ approaching $k$). This behavior is depicted in Fig. \ref{fig:PhaseSim15toEvanesc}. Here, again a 16 element array of $\lambda/2$ element spacing produces a beam at $15^\circ$ off broadside. The surface is now a linear phase surface with $k_s{=}0.95k$. It is obviously clear form the figure that the incident power on the surface is lost. 

\bibliography{Egorov_Eleftheriades_TAP2019_ForArxiv.bib}

\bibliographystyle{IEEEtran}

\begin{IEEEbiography}[{\includegraphics[width=1in,height=1.25in,clip,keepaspectratio]{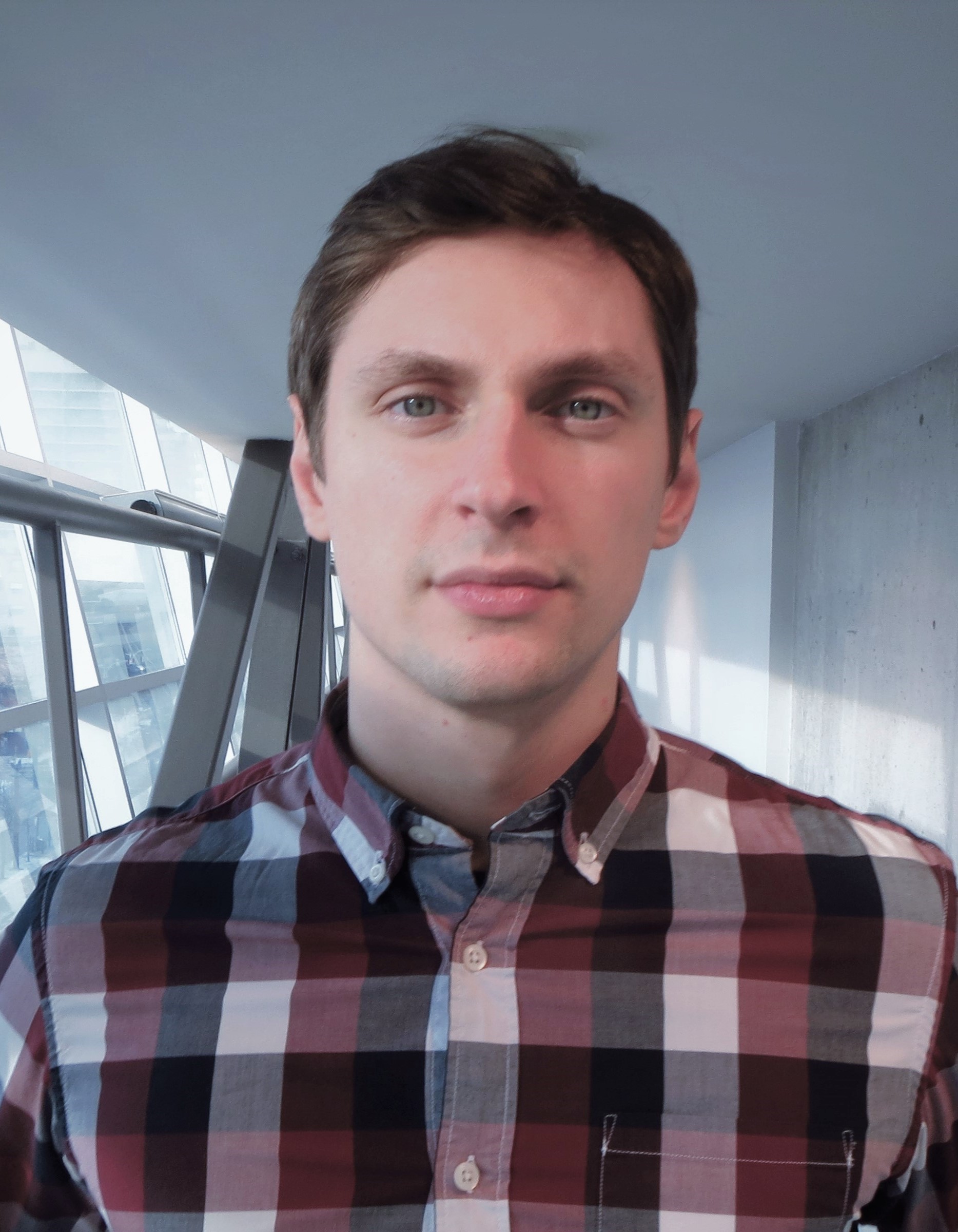}}]{Gleb A. Egorov} received the B.S. degree in electrical and computer engineering from the University of New Brunswick, Fredericton, Canada, in 2014, the M.A.Sc. degree in electrical engineering from the University of Toronto, Toronto, Canada, in 2016, and is currently working toward the Ph.D. degree in electrical engineering at the University of Toronto. 

His interests include theoretical physics, mathematical physics and philosophy of science and mathematics. 

Mr. Egorov was the recipient of the Governor General's Academic Bronze Medal (2010), the Lieutenant Governor of New Brunswick Silver Medal (2014) and the NSERC CGSM scholarship (2015) among other awards and scholarships.
\end{IEEEbiography}

\begin{IEEEbiography}[{\includegraphics[width=1in,height=1.25in,clip,keepaspectratio]{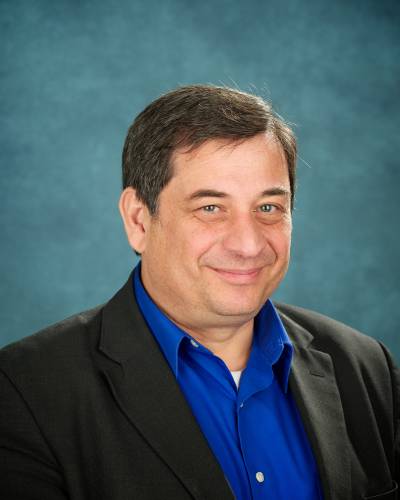}}]{George V. Eleftheriades} (S'86-M'88-SM'02-F'06)
received the M.S.E.E. and Ph.D. degrees in electrical
engineering from the University of Michigan, Ann
Arbor, MI, USA, in 1989 and 1993, respectively.

From 1994 to 1997, he was with the Swiss Federal
Institute of Technology, Lausanne, Switzerland. He
is currently a Professor with the Department of
Electrical and Computer Engineering, University of
Toronto, ON, Canada, where he holds the Canada
Research/Velma M. Rogers Graham Chair in Nanoand Micro-Structured Electromagnetic Materials. He
is a recognized international authority and pioneer in the area of metamaterials.
These are man-made materials which have electromagnetic properties not
found in nature. He introduced a method for synthesizing metamaterials using
loaded transmission lines. Together with his graduate students, he provided
the first experimental evidence of imaging beyond the diffraction limit and
pioneered several novel antennas and microwave components using these
transmission-line based metamaterials. His research has impacted the field by
demonstrating the unique electromagnetic properties of metamaterials; used
in lenses, antennas, and other microwave and optical components to drive
innovation in fields, such as wireless and satellite communications, defence,
medical imaging, microscopy, and automotive radar. He is currently leading
a group of graduate students and researchers in the areas of electromagnetic
and optical metamaterials, and metasurfaces, antennas and components for
broadband wireless communications, novel antenna beam-steering techniques,
far-field super-resolution imaging, radars, plasmonic and nanoscale optical
components, and fundamental electromagnetic theory.

Dr. Eleftheriades served as a member of the IEEE Antennas and Propagation
(AP)-Society administrative committee from 2007 to 2012. In 2009, he was
elected as a fellow of the Royal Society of Canada. He was an IEEE AP-S
Distinguished Lecturer from 2004 to 2009. His co-authored papers received
numerous awards such as the 2009 Best Paper Award from the IEEE MICROWAVE AND WIRELESS PROPAGATION LETTERS, the R. W. P. King
Best Paper Award from the IEEE TRANSACTIONS ON ANTENNAS AND
PROPAGATION in 2008 and 2012, and the 2014 Piergiorgio Uslenghi Best
Paper Award from the IEEE ANTENNAS AND WIRELESS PROPAGATION
LETTERS. He received the Ontario Premier’s Research Excellence Award
in 2001, the University of Toronto’s Gordon Slemon Award in 2001, and
the E.W.R. Steacie Fellowship from the Natural Sciences and Engineering
Research Council of Canada in 2004. He was a recipient of the 2008 IEEE
Kiyo Tomiyasu Technical Field Award and the 2015 IEEE John Kraus Antenna
Award, the 2019 IEEE AP-S Distinguished Achievement Award, and the
Research Leader Award from the Faculty of Applied Science and Engineering
of the University of Toronto in 2018. He served as the General Chair of
the 2010 IEEE International Symposium on Antennas and Propagation held
in Toronto, ON, Canada. He served as an Associate Editor for the IEEE
TRANSACTIONS ON ANTENNAS AND PROPAGATION.
\end{IEEEbiography}

% that's all folks
\end{document}